\documentstyle[aps]{revtex}

\begin{document}
\title{Renormalization and Short Distance Singular Structure.}
\author{Mario Castagnino}
\address{Instituto de Astronom\'{\i}a y F\'{\i}sica del Espacio.\\
Casilla de Correos 67, Sucursal 28.\\
1428 Buenos Aires, Argentina.}
\maketitle

\begin{abstract}
The relation between renormalization and short distance singular
divergencies in quantum field theory is studied. As a consequence a finite
theory is presented. It is shown that these divergencies are originated by
the multiplication of distributions (and worse defined mathematical
objects). Some of them are eliminated defining a multiplication based in
dimensional regularization while others disappear considering the states as
functionals over the observables space. Non renormalizable theories turn to
be finite, but anyhow they are endowed with infinite arbitrary constants.
\end{abstract}

\section{Introduction.}

Quantum Field Theory can be reduced to the knowledge of Wightman functions
(or T-ordered Feynman functions or retarded functions or euclidean
functions, etc.)\cite{Roman}, \cite{Haag}. These functions are short
distance singular mathematical objects (i. e. they diverge in the so called
''coincidence limits'' namely when some of their variables coincide) , e.
g.: the symmetric part of the two points Wightman function has a Hadamard
singularity, precisely: 
\begin{equation}
w^{(2)}(x,x^{\prime })=u\sigma ^{-1}+v\ln |\sigma |+w  \label{1.1}
\end{equation}
where $\sigma =\frac{1}{2}(x-x^{\prime })^{2}$, and $u,$ $v,$ $w$ are smooth
functions \footnote{%
For Wightman functions see \cite{Haag}, cap. VII, eq. (3.11). For Feynman
functions see \cite{De Witt} eqs. (17.61) and (16.72). For symmetrical
functions see \cite{Castagnino et al.}.}. These local singularities
originate the infinite ultraviolet results of Quantum Field Theory \cite
{Brown}\footnote{%
There is also another kind of potentially dangerous singularities as we will
see in section VI.}. To eliminate these infinities the theory must be
renormalized in such a way that meaningless divergent expressions become
meaningful. This technique is well known but not completely satisfactory
because by using it ''...we learned to peacefully coexist with alarming
divergencies... but these infinities are still with us, even though deeply
buried in the formalism.'' \cite{Roman}. On the other hand, as we know that
the short distance singularities are the cause of renormalization, if we
somehow remove these singularities we will directly obtain a finite and
exact Quantum Field Theory from the scratch. Phrased in another way: in this
paper we will find the short distance singularity in two quantum field
theory models and we will show that if these singularities are substracted
the theory turns out to be finite. The substraction of short distance
singularities has being essentially used for many years, e. g., in Quantum
Field Theory in Curved Space-Time, \cite{Castagnino et al.}, \cite{B&D}, 
\cite{Carmen} (and other chapters of quantum field theory e. g. \cite{Brown}
chapter 5), but it was not considered as a general method with a rational
motivation, as we are now trying to prove.

We hope that the study of the singular short distance structure will lead
us, in the future, either to find lagragians free of this sickness (may be
superstring or membrane lagrangians) or to find more elaborated ways to
remove this structure. Moreover, since the quantum field theory equations
can be highly not linear it will be clear that, in a general case, the
singular structure cannot be just removed by adding similar terms to those
of the bare lagrangian. The mechanism must be more general. Here we are
presenting the physical basis of this mechanism. Essentially we believe
that, since the origin of the problem are the short distance singularities,
philosophically it is wrong to put the blame on the old good lagrangian and
to torture it until it yields a finite theory. The cure must be provided
where the sickness is located\footnote{%
Renormalization would be like an electroshock. It works but we do not know
why it is so. We are looking for something like brain surgery, where the
disease is cured in the place where it is located.}.

We will find the singular structure using usual dimensional regularization 
\cite{Bocha} and, in the cases where it is possible, Hadamard regularization 
\cite{Castagnino et al.} and we will removed it by {\it two different ways}
at two different level of comprehension that we will discuss below.

\subsection{Simple substraction method. Detection of the local singularities.
}

In sections II to V we will review this well known method with three
purposes. i.- To introduce the main equations. ii.- To detect the local
singularities (as in eqs. (\ref{2.16}), (\ref{3.5}), (\ref{3.22}), (\ref
{4.2.7}), and (\ref{5.3})). iii.- To show the modification of the roles
played by the coupling constant when we go from the usual method to the new
one and to obtain renormalization group equations with the new method. We
will study the theory in a space of dimensions $n$. Generically the theory
will be finite for $n\neq 4$, but it will present short range singularities
for $n\rightarrow 4.$ E. g.: any two point function will have the structure: 
\begin{equation}
w^{(2)}(x-x^{\prime })=w^{(2)(s)}(x-x^{\prime })+w^{(2)(r)}(x-x^{\prime })
\label{1.2}
\end{equation}
where $w^{(2)(s)}(x-x^{\prime })$ is the singular component (in a sense that
we will precise below), that diverges when $n\rightarrow 4$ or $x\rightarrow
x^{\prime },$ and $w^{(2)(r)}(x-x^{\prime })$ is the regular one. The
substraction method, for these functions, is just to make the singular part
equal to zero or to subtract the singular part from $w^{(2)}(x-x^{\prime }).$
We will give two examples of this procedure:

i.-Scalar quantum field theory in a curved space-time (a theory invariant
under the group of general coordinates transformations, with no self
interaction and therefore with linear equations with variable coefficients)
in section II. In this case we only need two points functions as those of
eq. (\ref{1.2}).

ii.-$\lambda \phi ^{4}$ theory (a theory invariant under the Poincar\'{e}
group with self interaction and therefore with non-linear equations with
constant coefficients) studied in sections III, IV, and V. In the second
example we will need N-point functions.

These example are chosen not only because they are the simplest but also
because the two theories are quite different and cover a large range of
phenomena\footnote{%
E. g. conformal or trace anomaly, conservation of the energy momentum
tensor, etc. in example i.}

Then, let us precise how we will define the singular and the regular
components in the general case of N-point functions, in complete agreement
with the usual procedures of dimensional regularization. If $%
w^{(N)}(x_{1},x_{2},...,x_{N})$ are some (symmetric) N-point functions (like
Feynman or Euclidean functions) we can define the corresponding functional
generator (\cite{Haag}, eq. (II.2.21), \cite{Brown} eq. (3.2.11)) as: 
\begin{equation}
Z[\rho ]=\exp i\left\{ \frac{1}{N!}\sum_{N=0}^{\infty }\int
w^{(N)}(x_{1},x_{2},...,x_{N})\rho (x_{1})\rho (x_{2})...\rho
(x_{N})dx_{1}dx_{2}...dx_{N}\right\}  \label{1.3}
\end{equation}
where \footnote{%
The symbol $\sim $ means that the r.h.s. of the next equation can also be
truncated (\cite{Haag}, eqs. (II.2.18) and (II.2.23).}: 
\begin{equation}
w^{(N)}(x_{1},x_{2},...,x_{N})\sim \langle 0|\phi (x_{1})\phi (x_{2})...\phi
(x_{N})|0\rangle  \label{1.4}
\end{equation}
But, in a realistic field theory (namely a theory with interaction) these
functions are badly defined (as the two-point function of eq. (\ref{1.2}))
since they are objects with mathematical properties that are {\it worse than
those of the distributions,} because if these objects were distributions all
the integrals: 
\[
\int w^{(N)}(x_{1},x_{2},...,x_{N})\rho (x_{1})\rho (x_{2})...\rho
(x_{N})dx_{1}dx_{2}...dx_{N} 
\]
would be well defined (if, e. g.: $\rho (x)\in {\cal S}$ the Schwarz space).
But this is not the case, as we will see, so $Z[\rho ]$ and its derivatives
are not well defined\footnote{%
Namely, axiom B of \cite{Haag}, page 58, is only valid for free theories,
since from this axiom and Schwartz ''nuclear theorem'' it is shown that (\ref
{1.4}) is a distribution. Moreover, not only it is necessary that $Z[\rho ]$
would be well defined but also its $\partial /\partial \rho -$derivatives.
So all $w^{(N)}(x_{1},x_{2},...,x_{N})$ must be well defined functions after
renormalization.}.

As we have already said in the case of quantum filed theory in curved space
time we only deal with two point functions. But for the $\lambda \phi ^{4}$%
-theory we will deal with the two, four and six point functions, in the
coincidence limit where some points go to 0 and some points go to an
arbitrary value $z,$ because these are the only relevant functions in the
perturbation expansion of this theory up to $\lambda ^{2}$ order. So we will
be only interested in defining the singular and regular parts of the
functions $w^{(2)}(x_{1},x_{2}),$ in the coincidence limit $x_{1}=x_{2}=0,$
function $w^{(4)}(x_{1},x_{2},x_{3},x_{4}),$ in the coincidence limit $%
x_{1}=x_{2}=0,$ $x_{3}=x_{4}=z,$ and function $%
w^{(6)}(x_{1},x_{2},x_{3},x_{4},x_{5},x_{6}),$ in the coincidence limit $%
x_{1}=x_{2}=x_{3}=0,$ $x_{4}=x_{5}=x_{6}=z.$ We will see that these
coincidence limits have the general form $[w^{(2)}(0)]^{\beta
}[w^{(2)}(z)]^{\alpha },$ namely the product of the power of an infinite
quantity multiplied by the power of a distribution (or a worse mathematical
object). In fact, these powers appear in the higher order point functions
(see \cite{Haag} eq. (II.2.18)). So we have two problems that we will solve
using dimensional regularization:

i.- To obtain the regular part of $w^{(2)}(0).$ It is an easy problem since
via dimensional regularization $w^{(2)}(0)$ reads:

\begin{equation}
w^{(2)}(0)=\sum_{\gamma =0}^C\frac{d^{(\gamma )}}{(n-4)^\gamma }
\label{1.5'}
\end{equation}
where $C$ is a natural number and $d^{(\gamma )}$ are some coefficients.
Then the singular and regular components will be defined as: 
\begin{equation}
\lbrack w^{(2)}(0)]^{(s)}=\sum_{\gamma =1}^C\frac{d^{(\gamma )}}{%
(n-4)^\gamma }  \label{1.6'}
\end{equation}
and 
\begin{equation}
\lbrack w^{(2)}(z)]^{(r)}=d^{(0)}  \label{1.7'}
\end{equation}

Then the regular part of $[w^{(2)}(0)]^\beta $ is simply $[d^{(0)}]^\beta .$

ii.- To obtain the regular part of $[w^{(2)}(z)]^{\alpha }$. This is a more
difficult problem since we must {\it multiply the ill defined function} $%
w^{(2)}(x_{1},x_{2})$ {\it by itself.} But function $w^{(2)}(x_{1},x_{2})$
is worse than a distribution, so it cannot be multiplied by itself in a
unique and well defined way\footnote{%
Here is where one type of the divergencies is ''deeply buried in the
formalism'' \cite{Roman}. We will find another type of potentially dangerous
divergencies in section VI.}. Thus we will be forced to define the
multiplication procedure for, e. g.: $[w^{(2)}]^{2}$ and $[w^{(2)}]^{3},$ in
an ad hoc way based on {\it dimensional regularization} (see \cite{Brown}
pages. 162 to 167 and 207 to 214). To stress this fact we will call them $%
[w^{(2)}]^{(d)2}$ and $[w^{(2)}]^{(d)3}$ respectively (where the susperindex
''d'' comes from ''dimensional regularization''). Then the {\it %
multiplication procedure} will be the following:

a.-Using dimensional regularization we will find that the powers are regular
when $n\neq 4,$ but when $n\rightarrow 4$ they behave as: 
\begin{equation}
\lbrack w^{(2)}(z)]^{(d)\alpha }=\sum_{\delta =0}^D\frac{d^{(\alpha ,\delta
)}(z)}{(n-4)^\delta }  \label{1.5}
\end{equation}
where $D$ is a natural number and $d^{(\alpha ,\delta )}(z)$ are
distributions (showing that, in effect, the objects we are dealing with are
worse than distributions).

b.- The singular and regular component will be defined as: 
\begin{equation}
\lbrack w^{(2)}(z)]^{(d)\alpha (s)}=\sum_{\delta =1}^D\frac{d^{(\alpha
,\delta )}(z)}{(n-4)^\delta }  \label{1.6}
\end{equation}
and 
\begin{equation}
\lbrack w^{(2)}(z)]^{(d)\alpha (r)}=d^{(\alpha ,0)}(z)  \label{1.7}
\end{equation}

Moreover, the multiplication (ii) and the procedure to take the regular part
for $z=0$ (i) are {\it not commutative}. After these definitions we can
substitute $[w^{(2)}(0)]^{\beta }$ and $[w^{(2)}(z)]^{\alpha }$ by $%
[w^{(2)}(0)^{(r)}]^{\beta }$ and $[w^{(2)}(z)]^{(d)\alpha (r)}.$ Then if we
consider only these regular parts, which are in general distributions (but
regular functions in the two examples below), the functional generator $%
Z[\rho ]$ and its derivatives (eq. (\ref{1.3})) turns out to be well defined
as well as the theory that it generates. The existence of singularities like
those of the above equations is proved by the examples below (see also
section V). The decompositions (\ref{1.6'}), (\ref{1.7'}) and (\ref{1.6}), (%
\ref{1.7}) are not unique, since $\infty =\infty +c$ or $\infty =c.\infty ,$
for any finite $c.$ This ambiguity will be present in our method, as in
ordinary renormalization theory, and it yields the running coupling
constants and the renormalization group, as we will see.

\subsection{Functional method.}

In section VI we will present a mathematical structure that naturally yields
the elimination of the singularities. We will follow the line of thought of
papers \cite{Laura} and \cite{Deco} where a formalism to deal with systems
with continuous spectrum was introduced. It proves to be useful in the study
of decaying, equilibrium, and decoherence (where we have defined a final
intrinsically consistent set of histories). So we claim that perhaps it is a 
{\it general formalism} that can also be used in the problem of this paper.
This mathematical structure would also be the rational justification of the
somehow dictatorial or childish substraction method. This is the main
contribution of the paper. The idea is the following: Coarse-graining is a
well known technique where some features of a system are considered relevant
while others are not\footnote{%
Or, in observables language, the observables of theory measure only the
relevant features.}. The functional method of papers \cite{Laura}, \cite
{Deco} is a generalization of coarse-graining \footnote{%
E. g.: classically coarse-graining is just the particular case where the
functionals are built using the characteristic functions of lattices in
phase space (see \cite{Laura}).}, where the states are considered as
functional over a certain space of observables\footnote{%
Moreover, this is the natural way to face the problem since the observables
are more primitive objects than the states \cite{Haag}.}. Using this
philosophy we will postulate that physical observables are such that cannot
see the singular components of the states, because these components are
irrelevant for these observables. Symmetrically, singularities could be
contained in the observables and we can postulate that physical states
cannot see the singular part of the observables\footnote{%
Really this will be the case since observables are products like $\phi
(x_{1})\phi (x_{2})...$of field $\phi (x),$ which are distributions or worse
defined mathematical objects.}. In this way we will obtain the automatic
substraction of all kinds of the singularities. There is a good physical
reason for this postulate: the singularities (either of states or
observables) are just mathematical artifacts originated in the
oversimplified lagrangian that we usually choose. Then, clearly physical
observables or states cannot see these mathematical unphysical objects. In a
more intuitive language: the physical observables or states do not see the
singularities because they are too small (point-like). Possibly the physical
observables and states just see up to Planck's length \footnote{%
We could as well postulate that the singular part of the observables see the
singular part of the state. Even if there are physical reasons to introduce
this postulate in the case of decoherence, this reasons are absent in the
case of renormalization (see section VI.B)}.

Using the Jaynes philosophy \cite{Jaynes} we can say that if {\it physical }%
observables can not see {\it mathematical} singularities (which in fact is a
very reasonable position) then the (singular) states of the usual theory are
really biased objects because they contain {\it arbitrary unphysical
information} (i. e. the singularities) that cannot be measured by the
physical apparatuses that we have in our laboratory i. e. our physical
observables (and really this is an experimental fact: since apparatuses
measure the values given by the finite renormalized theory). Then the (rough
material) singular states, observables, and the mean values obtained with
them are {\it biased over-informed} objects containing dubious information,
because in fact ''we have a basic ignorance of the nature of infinite
energies or infinitesimal distances'' (\cite{Brown}, page. 63), while
renormalized (or free of any kind of singularities) states, observables, and
mean values are {\it unbiased} objects containing just the physical
information available. In fact, to suppose that we know and measure
everything would be an ''inexcusable hubris'' (\cite{Brown}, page. 64).
Moreover the resulting theory turns out to be insensitive to our degree of
knowledge (originated in the more or less precision of our measurement
apparatuses), thus we simply{\it \ postulate that this degree of knowledge
is, and cannot be, infinite. }All this philosophy is embodied in the
mathematical structure studied in section VI.

We will discuss our conclusions in section VII.

\section{First method: scalar quantum field theory in curved space-time%
\protect\footnote{%
The expert reader may go directly to section V and consider sections II to
IV as a {\it didactical appendix}} to be read after section VII. But we
consider that this didactical discussion is essential in order to convince
the reader that the new formalism also works in practice..}

This theory is the simplest non-trivial example of the method, the theory of
a scalar neutral massive fields in a curved space-time (of dimension $n,$
since we need a formalism prepared for dimensional regularization) with
metric $g_{\mu \nu }(x)$. Let us consider the action \footnote{%
For the sake of conciseness we do not demonstrate the basic equations of
quantum field theory in curved space-time. We just quote the number of the
equation of reference \cite{B&D} at the beginning of each of them. In
sections III, IV, and V we will use reference \cite{Brown} for the same
purpose in the $\lambda \phi ^4$ case.}: 
\begin{equation}
(6.9)\qquad S=S_{g}+S_{m}  \label{2.1}
\end{equation}
where: 
\begin{equation}
(6.11)\qquad S_{g}=\int (-g)^{\frac{1}{2}}(16\pi G_{0})^{-1}(R-2\Lambda
_{0})d^{n}x  \label{2.2}
\end{equation}
and: 
\begin{equation}
S_{m}=\int (-g)^{\frac{1}{2}}L_{m}d^{n}x  \label{2.3}
\end{equation}
where $L_{m}$ is the matter lagrangian: 
\begin{equation}
(3.24)\qquad L_{m}(x)=\frac{1}{2}\left\{ g^{\mu \nu }(x)\phi _{,\mu }(x)\phi
_{,\nu }(x)-[m^{2}+\xi R(x)]\phi ^{2}\right\}  \label{2.4}
\end{equation}
$G_{0}$ and $\Lambda _{0}$ are the bare Newton and cosmological constants
respectively, $m$ is the scalar field mass, $g^{\mu \nu }$ the inverse
metric tensor (signature +,-,-,-), $g$ its determinant, $\xi $ a numerical
factor, and $R(x)$ the Ricci scalar. For an in-out scattering we can define
the functional generator $Z[\rho ]$ such that: 
\begin{equation}
(6.15)\qquad Z[0]=\langle out,0|in,0\rangle =e^{iW}  \label{2.5}
\end{equation}
so: 
\begin{equation}
(6.19)\qquad W=-i\ln \langle out,0|in,0\rangle  \label{2.5'}
\end{equation}
Then $W$ can be computed using the effective lagrangian $L_{eff}$, defined
by: 
\begin{equation}
(6.36)\qquad W=\int [-g(x)]^{\frac{1}{2}}L_{eff}(x)d^{n}(x)  \label{2.6}
\end{equation}
where $L_{eff}$ reads: 
\begin{equation}
(6.37)\qquad L_{eff}(x)=\frac{i}{2}\lim_{x^{\prime }\rightarrow
x}\int_{m^{2}}^{\infty }dm^{2}\Delta _{F}^{DS}(x,x^{\prime })  \label{2.7}
\end{equation}
where $\Delta _{F}^{DS}(x,x^{\prime })$ is the
De-Witt-Schwinger-Feynman-Green function: 
\[
(3.138)\qquad \Delta _{F}^{DS}(x,x^{\prime })=-i\Delta ^{\frac{1}{2}%
}(x,x^{\prime })(4\pi )^{-\frac{n}{2}}\times 
\]
\begin{equation}
\int_{0}^{\infty }ids(is)^{-\frac{n}{2}}\exp [-im^{2}s+\frac{\sigma }{2is}%
]F(x,x^{\prime };is)  \label{2.8}
\end{equation}
The $\sigma (x,x^{\prime })$ is half the square of the geodesic distance
between $x$ and $x^{\prime }$, $\Delta (x,x^{\prime })$ is the van
Vleck-Morette determinant, and 
\begin{equation}
(3.137)\qquad F(x,x^{\prime };is)=a_{0}(x,x^{\prime })+a_{1}(x,x^{\prime
})is+a_{2}(x,x^{\prime })(is)^{2}+...  \label{2.9}
\end{equation}
where the $a$ coefficients can be obtained from ref. \cite{B&D} eqs. (3.131,
2, 3) and corresponds to an expansion in the metric $g_{\mu \nu }(x)$ and
its derivatives, precisely to orders 0, 2, 4,...in these derivatives. The
coefficients are biscalars, namely all the formalism is covariant under
general coordinates transformation.

Eq. (\ref{2.7}) is the simple non trivial example of the relation between $%
L_{eff}$ and the two points function $\Delta _F^{DS}(x,x^{\prime })$ in the
limit $x\rightarrow x^{\prime }$, where in fact $\Delta ^{DS}(x,x^{\prime })$
has a short distance singularity that makes $L_{eff}$ a divergent quantity
as we will see. If we want to retain the $n=4$ dimension of $L_{eff}$ as $%
(lenght)^{-4}$ when $n\neq 4,$ we must introduce an arbitrary mass $\mu .$
Then $L_{eff}$ reads: 
\begin{equation}
(6.42)\qquad L_{eff}=\frac 12(4\pi )^{-\frac n2}\left( \frac m\mu \right)
^{n-4}\sum_{j=0}^\infty a_j(x)m^{4-2j}\Gamma \left( j-\frac n2\right)
\label{2.10}
\end{equation}
where $a_j(x)=a_j(x,x)$ are functions of the curvatures and its derivatives,
and the $\Gamma $ function diverges when $n\rightarrow 4$.

\subsection{Renormalization using dimensional regularization}

By the dimensional regularization method everything is now prepared to
renormalize the theory. When $n\rightarrow 4$ the first three terms (those
that correspond to orders 0, 2, 4) diverge and we obtain the divergent or
singular component of $L_{eff}$ that reads (we have dropped the $O(n-4)$
terms): 
\[
(6.44)\qquad L^{(s)}(x)=-(4\pi )^{-\frac{n}{2}}\left\{ \frac{1}{n-4}+\frac{1%
}{2}\left[ \gamma +\ln \left( \frac{m^{2}}{\mu ^{2}}\right) \right] \right\}
\times 
\]
\begin{equation}
\left( \frac{4m^{4}a_{0}(x)}{n(n-2)}-\frac{2m^{2}a_{1}(x)}{n-2}%
+a_{2}(x)\right)  \label{2.11}
\end{equation}
where: 
\[
(6.46)\qquad a_{0}(x)=1 
\]
\[
(6.47)\qquad a_{1}(x)=\left( \frac{1}{6}-\xi \right) R 
\]
\begin{equation}
(6.48)\qquad a_{2}(x)=\frac{1}{180}R_{\alpha \beta \gamma \delta }R^{\alpha
\beta \gamma \delta }-\frac{1}{180}R_{\alpha \beta }R^{\alpha \beta }-\frac{1%
}{6}\left( \frac{1}{5}-\xi \right) \Box R+\frac{1}{2}\left( \frac{1}{6}-\xi
\right) ^{2}R^{2}  \label{2.11'}
\end{equation}
where $R_{\alpha \beta \gamma \delta }$ is the curvature tensor and $%
R_{\alpha \beta }=R_{\text{ }\alpha \mu \beta }^{\mu }$. The usual
renormalization procedure is to absorb this singular component in the bare $%
S_{g},$ so we can renormalize $G_{0}$ and $\Lambda _{0}$ as: 
\begin{equation}
(6.50)\qquad \Lambda _{phys}=\Lambda _{0}+\frac{32\pi m^{2}G_{0}}{(4\pi )^{%
\frac{n}{2}}n(n-2)}\left\{ \frac{1}{n-4}+\frac{1}{2}\left[ \gamma +\ln
\left( \frac{m^{2}}{\mu ^{2}}\right) \right] \right\}  \label{2.12}
\end{equation}
\begin{equation}
(6.51)\qquad G_{phys}=G_{0}/1+16G_{0}\frac{2m^{2}\left( \frac{1}{6}-\xi
\right) }{(4\pi )^{\frac{n}{2}}(n-2)}\left\{ \frac{1}{n-4}+\frac{1}{2}\left[
\gamma +\ln \left( \frac{m^{2}}{\mu ^{2}}\right) \right] \right\}
\label{2.13}
\end{equation}
(where we have neglected the squares terms in the bare constants) so we
choose $G_{0}$ and $\Lambda _{0}$ in such a way that $G_{phys}$ and $\Lambda
_{phys}$ turn out to be finite when $n=4.$ But this is not enough since the
divergence of the $a_{2}(x)$ term cannot be eliminated in this way, so the
theory with action $S_{g}$ is not renormalizable. But, if we add three $"H"$
terms to the gravitational lagrangian which are linear combinations of the
three terms of eq. (\ref{2.11'}): i. e., linear combinations of $R^{2},$ $%
R_{\mu \nu }R^{\mu \nu },$ $R_{\alpha \beta \gamma \delta }R^{\alpha \beta
\gamma \delta },$ and $\Box R,$ (there are only three $"H"$ terms because
there is a relation among the last four terms) and renormalize the three
corresponding coefficients (known as $\alpha ,$ $\beta ,$ $\gamma )$ the
theory becomes renormalizable and finite (see \cite{B&D} eqs. (6.52) to
(6.56)). So from now on we will consider that these $"H"$ terms are added to
the gravitational lagrangian (\ref{2.2}).

But let us observe that essentially what we have done with this standard
renormalization recipe is to define, as proved in ref. \cite{B&D}, a
regular-substracted lagrangian \footnote{%
We are using a particular criterion to define the singular component. This
criterion is neither unique not irrelevant \cite{JuanPa}. It is clear that
the singular term must have the form $\infty \times geometrical$ $object$
(namely invariant under general coordinates transformations) But this object
can be chosen in a variety of ways, since, as we have said, we know that $%
\infty =\infty +c$ or $\infty =c.\infty $ for any finite $c.$}, that for $%
n=4 $ reads: 
\begin{equation}
(6.59)\qquad L^{(r)}=L_{eff}-L^{(s)}=\frac{1}{32\pi ^{2}}\int_{0}^{\infty
}\sum_{j=3}^{\infty }a_{j}(x)(is)^{j-3}e^{-im^{2}s}ids  \label{2.14}
\end{equation}
which turns out to be finite and can be used instead of the divergent $%
L_{eff}$ \footnote{%
Eq. (\ref{2.14}) shows that already in ref. \cite{B&D} substraction was used
in quantum field theory in curved space-time, as we have said in the
introduction.}$.$ Thus we can foresee that both the standard renormalization
recipe and the substraction recipe coincide. What we have really made is a
substraction using dimensional regularization. Making the same substraction
in $\Delta _{F}^{DS}(x,x^{\prime })$ (eq. (\ref{2.9})) we will obtain the
regular $\Delta _{F}^{DS(r)}(x,x^{\prime })$ \footnote{%
Since this is the only non vanishing truncated point function in the theory,
all ordinary point functions of the theory are finite and they can be
directly computed.}. We will make this calculation in the next section using
Hadamard regularization \cite{Castagnino et al.} because using this method
we can better show the presence and nature of the local singularities.

\subsection{Hadamard regularization and the substraction recipe.}

Let us now see how we can directly work in the $n=4$ case. The divergencies
now appear when $x\rightarrow x^{\prime }$ (not when $n\rightarrow 4$ as in
the previous section). In this section we will see how the two singular
behaviors are related. The effective lagrangian (\ref{2.10}) reads: 
\[
(6.38)\qquad L_{eff}=-\lim_{x^{\prime }\rightarrow x}\frac{\Delta
^{1/2}(x,x^{\prime })}{32\pi ^{2}}\int_{0}^{\infty }\frac{ds}{s^{3}}%
e^{-(m^{2}s-\sigma /2s)}\times 
\]
\begin{equation}
\left[ a_{0}(x,x^{\prime })+a_{1}(x,x^{\prime })is+a_{2}(x,x^{\prime
})(is)^{2}+...\right]  \label{2.15}
\end{equation}
From eqs. (\ref{2.8}) we may compute: 
\[
\Delta _{F}^{DS}(x,x^{\prime })=-i\frac{\Delta ^{1/2}(x,x^{\prime })}{(4\pi
)^{2}}\int_{0}^{\infty }ids(is)^{-2}e^{-(m^{2}s-\sigma /2s)}\times 
\]
\begin{equation}
\left[ a_{0}(x,x^{\prime })+a_{1}(x,x^{\prime })is+a_{2}(x,x^{\prime
})(is)^{2}+...\right] =\overline{\Delta _{F}^{DS}(x,x^{\prime })}+\frac{1}{2}%
i\Delta _{F}^{DS(1)}(x,x^{\prime })  \label{2.16}
\end{equation}
where ((\cite{De Witt}, eqs. (17.61), (17.62)): 
\[
\overline{\Delta _{F}^{DS}(x,x^{\prime })}=\frac{\Delta ^{1/2}a_{0}}{8\pi }%
\delta (\sigma )-\frac{\Delta ^{1/2}}{8\pi }\theta (\sigma )\left[ \frac{1}{2%
}(m^{2}a_{0}-a_{1})-\frac{2\sigma }{2^{2}.4}(m^{4}a_{0}-2m^{2}a_{1}+2a_{2})%
\right. 
\]
\begin{equation}
\left. +\frac{(2\sigma )^{2}}{2^{2}.4^{2}.6}%
(m^{6}a_{0}-3m^{4}a_{1}+6m^{2}a_{2}-6a_{3})+...\right]  \label{2.16'}
\end{equation}
and 
\[
\Delta _{F}^{DS(1)}(x,x^{\prime })=-\frac{\Delta ^{1/2}a_{0}}{4\pi
^{2}\sigma }+\frac{\Delta ^{1/2}}{2\pi ^{2}}\log \frac{e^{\gamma }}{2}%
|2m^{2}\sigma |\left[ \frac{1}{2}(m^{2}a_{0}-a_{1})\right. 
\]
\[
\left. -\frac{2\sigma }{2^{2}.4}(m^{4}a_{0}-2m^{2}a_{1}+a_{2}\right] 
\]
\[
-\frac{\Delta ^{1/2}}{2\pi ^{2}}\left[ \frac{1}{4}m^{2}a_{0}-\frac{2\sigma }{%
2^{2}.4}\left( \frac{5}{4}m^{4}-2m^{2}a_{1}-a_{2}\right) \right. 
\]
\[
\left. +\frac{(2\sigma )^{2}}{2^{2}.4^{2}.6}\left( \frac{5}{3}m^{6}a_{0}-%
\frac{9}{2}m^{4}a_{1}+\frac{15}{2}m^{2}a_{2}-\frac{9}{2}a_{3}\right)
+...\right] 
\]
\[
\frac{\Delta ^{1/2}}{2\pi ^{2}}\left[ \left( \frac{a_{2}}{4m^{2}}+\frac{a_{3}%
}{4m^{4}}+\frac{a_{4}}{8m^{6}}+...\right) \right. 
\]
\begin{equation}
\left. -\frac{2\sigma }{2^{2}.4}\left( \frac{a_{3}}{m^{2}}+\frac{a_{4}}{m^{4}%
}+...\right) +...\right]  \label{2.16''}
\end{equation}
According to dimensional regularization the singular part of $\Delta
_{F}^{DS}(x,x^{\prime })$ corresponds to the one with coefficients $%
a_{0},a_{1,}a_{2}$ (see (\ref{2.11})). The remaining terms are the regular
part (see (\ref{2.14})). Then:

\[
\Delta _{F}^{DS(s)}(x,x^{\prime })=\frac{\Delta ^{1/2}a_{0}}{8\pi }\delta
(\sigma )-\frac{\Delta ^{1/2}}{8\pi }\theta (\sigma )\left[ \frac{1}{2}%
(m^{2}a_{0}-a_{1})-\frac{2\sigma }{2^{2}.4}(m^{4}a_{0}-2m^{2}a_{1}+2a_{2})%
\right] + 
\]
\[
+\frac{i}{2}\left\{ \frac{\Delta ^{1/2}a_{0}}{4\pi ^{2}\sigma }+\frac{\Delta
^{1/2}}{2\pi ^{2}}\log \frac{e^{\gamma }}{2}|2m^{2}\sigma |\left[ \frac{1}{2}%
(m^{2}a_{0}-a_{1})\right. \right. 
\]
\[
\left. -\frac{2\sigma }{2^{2}.4}(m^{4}a_{0}-2m^{2}a_{1}+a_{2}\right] 
\]
\[
-\frac{\Delta ^{1/2}}{2\pi ^{2}}\left[ \frac{1}{4}m^{2}a_{0}-\frac{2\sigma }{%
2^{2}.4}\left( \frac{5}{4}m^{4}-2m^{2}a_{1}-a_{2}\right) \right. 
\]
\begin{equation}
\left. \left. +\frac{(2\sigma )^{2}}{2^{2}.4^{2}.6}\left( \frac{5}{3}%
m^{6}a_{0}-\frac{9}{2}m^{4}a_{1}+\frac{15}{2}m^{2}a_{2}\right) +...\right] +%
\frac{\Delta ^{1/2}}{2\pi ^{2}}\frac{a_{2}}{4m^{2}}\right\}  \label{2.17}
\end{equation}
This $\Delta _{F}^{DS(s)}(x,x^{\prime })$ contains all the terms that
diverges when $\sigma \rightarrow 0$ (like $\delta (\sigma ),1/\sigma ,\log
\sigma )$ plus the terms with a divergent first derivative when $\sigma
\rightarrow 0$ (like $\theta (\sigma ),$ $\sigma \theta (\sigma ),\sigma
\log \sigma )$ plus some convergent terms when $\sigma \rightarrow 0$ (like $%
1,$ $\sigma ,$ $\sigma ^{2}).$ In this way we arrive to the first important
conclusion of this section: {\it The poles of }$\Gamma \left( j-\frac{n}{4}%
\right) ${\it \ which originate the three coefficients }$a_{0},a_{1,}a_{2}$ 
{\it correspond to the divergent terms or the terms with divergent
derivative when }$\sigma \rightarrow 0.$ There also are convergent terms in $%
\Delta _{F}^{DS(s)}(x,x^{\prime })$ but they are physically irrelevant as we
will soon see. The regular part of $\Delta _{F}^{DS}(x,x^{\prime })$ reads 
\[
\Delta _{F}^{DS(r)}(x,x^{\prime })=-\frac{\Delta ^{1/2}}{8\pi }\theta
(\sigma )\left[ \frac{(2\sigma )^{2}}{2^{2}.4^{2}.6}(-6a_{3})+...\right] +%
\frac{i}{2}\left\{ \frac{\Delta ^{1/2}}{2\pi ^{2}}\log \frac{e^{\gamma }}{2}%
|2m^{2}\sigma |[\sigma ^{2}+...\right. 
\]
\begin{equation}
-\frac{\Delta ^{1/2}}{2\pi ^{2}}\left[ \frac{(2\sigma )^{2}}{2^{2}.4^{2}.6}%
\left( -\frac{9}{2}a_{3}\right) \right] +\frac{\Delta ^{1/2}}{2\pi ^{2}}%
\left[ \left( \frac{a_{3}}{4m^{4}}+\frac{a_{4}}{8m^{6}}+...\right) -\frac{%
2\sigma }{2^{2}.4}\left( \frac{a_{3}}{m^{2}}+\frac{a_{4}}{m^{4}}+...\right)
+\right]  \label{2.17'}
\end{equation}
and contains terms that are convergent and with first derivative also
convergent when $\sigma \rightarrow 0.$

Then we can define the ''Hadamard regularization'' as the prescription that
the singular part of $\Delta _{F}^{DS}(x,x^{\prime })$ contains all the
terms divergent or with first derivative divergent when $\sigma \rightarrow
0 $ while the regular part of $\Delta _{F}^{DS(r)}(x,x^{\prime })$ contains
the terms which are convergent and with convergent first derivative when $%
\sigma \rightarrow 0.$ At first sight the dimensional regularization and the
Hadamard regularization do not coincide since in $\Delta
_{F}^{DS(s)}(x,x^{\prime })$ there are convergent terms with all their
derivatives, namely those like $1,\sigma ,\sigma ^{2}.$ Nevertheless the
difference is physically irrelevant since these terms are multiplied by $%
a_{0}(x,x^{\prime }),a_{1}(x,x^{\prime }),a_{2}(x,x^{\prime })$ that when $%
\sigma \rightarrow 0$ have the limits: 
\begin{equation}
\lim_{x^{\prime }\rightarrow x}a_{i}(x,x^{\prime })=a_{i}(x),\qquad i=1,2,3
\label{2.18}
\end{equation}
From eq. (\ref{2.11'}) we see that these terms are proportional to the
linear combinations of $I,$ $R,$ $R^{2},$ $R_{\mu \nu }R^{\mu \nu },$ $%
R_{\alpha \beta \gamma \delta }R^{\alpha \beta \gamma \delta },$ and $\Box
R, $ contained in the terms of the gravitational lagrangian. Therefore terms
we are discussing can be absorbed in the gravitational action $S_{g}$
supplemented by the $H$ terms. Then to unify the two regularizations $\Delta
_{F}^{DS(r)}(x,x^{\prime })$ must be defined modulo some terms with
arbitrary coefficients corresponding to the undefined terms $1,\sigma
,\sigma ^{2}$ . In the effective lagrangian these terms will produce finite
terms that can be added to $\Lambda ,G,\alpha ,\beta ,\gamma $ (\cite{B&D}
eq.(6.60)). The coefficients of these terms will be called $l,g,a,b,c.$
Dropping these terms for the moment, we can compute the regular lagrangian
corresponding to $\Delta _{F}^{DS(r)}(x,x^{\prime })$ that in the
coincidence limit reads:

\begin{equation}
\lim_{x\rightarrow x^{\prime }}\Delta _F^{DS(r)}(x,x^{\prime
})=\lim_{x^{\prime }\rightarrow x}\frac{i\Delta ^{\frac 12}}{4\pi ^2}\left[ 
\frac{a_3}{4m^4}+\frac{a_4}{8m^6}+...\right]  \label{2.19}
\end{equation}

Then as $\lim_{x^{\prime }\rightarrow x}\Delta =1$ (\cite{De Witt}, eq.
(17.86)) we have that: 
\begin{equation}
\lim_{x\rightarrow x^{\prime }}\Delta _{F}^{DS(r)}(x,x^{\prime })=\frac{i}{%
4\pi ^{2}}\left\{ \frac{a_{3}}{4m^{4}}+\frac{a_{4}}{8m^{6}}+...\right\}
\label{2.21}
\end{equation}
We may now add the arbitrary coefficient terms and we obtain 
\begin{equation}
\lim_{x\rightarrow x^{\prime }}\Delta _{F}^{DS(r)}=\frac{i}{(4\pi )^{2}}%
\left\{ 4lm^{2}+ga_{1}+\frac{a_{2}}{m^{2}}+\frac{a_{3}}{m^{4}}+...\right\}
\label{2.22}
\end{equation}
where $l$ and $g$ are the already defined arbitrary coefficients and those $%
a,b,c,$ corresponding to $\alpha ,\beta ,\gamma $ are hidden in $a_{2}$.
Using eq. (\ref{2.7}) we obtain \footnote{%
In the first two terms , instead of $\int_{m^2}^\infty $ we use $%
-\int_0^{m^2}$ and in the third term $-\int_1^{m^2},$ because they work in
these terms as $\int_{m^2}^\infty $ in the rest of the terms (see \cite{B&D}%
, pag. 157).}: 
\begin{equation}
L^{(r)}(x)=\frac{1}{32\pi ^{2}}\left[ 2lm^{4}+gm^{2}a_{1}+a_{2}\log m^{2}+%
\frac{a_{3}}{m^{2}}+...\right]  \label{2.19x}
\end{equation}
which turns out to be equal to eq. (\ref{2.14}) (except that in the quoted
equation the three first terms are missing since they are absorbed in $S_{g}$
supplemented by the $"H"$ terms), showing the coincident of the two methods.

Therefore the substracted $S^{(r)}$ reads: 
\begin{equation}
S^{(r)}=\int (-g)^{\frac 12}\left[ -\frac{2\Lambda _0}{16\pi ^2G_0}+\frac{%
m^4l}{16\pi ^2}+\frac R{16\pi ^2G_0}+\frac 16\frac{gm^2R}{32\pi ^2}+\frac{%
\log m^2a_2}{32\pi ^2}+\frac{a_3}{32\pi ^2m^2}+...\right]  \label{2.20}
\end{equation}
where the quantities $-\frac{2\Lambda _0}{16\pi ^2G_0}+\frac{m^4l}{16\pi ^2}$
and $\frac 1{16\pi ^2G_0}+\frac 16\frac{gm^2}{32\pi ^2}$ must be determined
by physical measurements (as the $\alpha ,\beta ,\gamma $ that are hidden in 
$a_2).$

So using Hadamard regularization and the substraction recipe the result is,
somehow, simpler since eqs. (\ref{2.12}) and (\ref{2.13}) just read. 
\begin{equation}
G_{phys}=G_{0}/1+\frac{1}{6}G_{0}gm^{2},\qquad \Lambda _{phys}=\Lambda _{0}-%
\frac{1}{2}G_{0}m^{2}l  \label{2.21'}
\end{equation}
so the bare constants are finite and would coincide, from the very
beginning, with the physical ones for the choice $l=g=0$ of the arbitrary
coefficients $l$ and $g${\it . }Thus using Hadamard regularization and the
substraction recipe: ''{\it we must remove }$\Delta ^{(s)}${\it \ from }$%
\Delta ${\it \ and use }$\Delta ^{(r)}"$ we have obtained the same result of
section II.A: all the infinities are removed and substituted by finite
quantities. Thus ''substraction recipe'' works as the standard
renormalization. The new recipe just consist in the elimination of the
singular (or with singular first derivative) short distance components of
the two point function $\Delta _{F}^{DS}(x,x^{\prime })$, the only relevant
truncated two point function in this theory. If we would have a $\lambda
\phi ^{4}$ interaction more truncated point functions must be substracted,
as we will see in the next example.

\section{First method. $\lambda \phi ^4$ theory in the lowest order.}

In this section we will use the substraction method in the $\lambda \phi
^{4} $ theory with lagrangian\footnote{%
In sections III, IV, and V the numbers before the equations correspond to
ref. \cite{Brown}. Moreover, comparing eqs. (\ref{2.4}) with (\ref{3.0}) we
see that there is a change of convention in the sign of the norm, so in the
following sections we change this convention in order that our equations
would coincide with those of the corresponding references. Also, in order to
comply with ref. \cite{Brown} we will use, sometimes, $\Delta _{F}(x)$ and,
sometimes, $\Delta _{E}(x).$}: 
\begin{equation}
(3.3.1)\qquad L=-\frac{1}{2}(\partial _{\mu }\phi )^{2}-\frac{1}{2}m^{2}\phi
^{2}-\frac{1}{4!}\lambda \phi ^{4}+\Lambda  \label{3.0}
\end{equation}
Dimensional regularization and minimal substraction will be done following
ref. \cite{Brown}.

It must be clear that, as we will isolate the divergent parts and then
substract them, the theory will necessarily turns out to be finite. Thus our
only aim, in sections III, IV, and V, is to detect the local divergencies
and to compare our method with the usual one to see how the results are
obtained and to show that they are similar, (so in each paragraph ''i'' we
will see how we can find the singular and regular parts of the objects
appearing in the theory, in ''ii'' we will review usual renormalization but
using our notation, and in ''iii'' we will see how substraction recipe
handles the divergence problem and compare the results)

\subsection{Singular and regular parts of $\Delta _E(0)$ and mass
renormalization.}

i.- From eq. (\ref{1.1}) we know that $\Delta _{E}(x)$ is one of the main
characters of the play. It is divergent when $x\rightarrow 0.$ So we will
define the singular and the regular parts of $\Delta _{E}(0),$ first using
dimensional diagonalization and then the Hadamard one \footnote{%
In both cases the singular component will have the form $\infty \times
geometrical$ $object$ (in this case invariant under a Lorentz
transformation). Of course there are many possible substractions, as in the
previous section. In section III.B we will pick the minimal one as in ref. 
\cite{Brown}. In section III.C the Hadamard one and we will show the finite
difference between the two choices.}. In $n$ dimensions it reads (just
computing the tadpole graph and neglecting non connected graphs that will be
taken into account in III.B and IV.B) :

\begin{equation}
(4.3.8)\qquad \lim_{x\rightarrow x^{\prime }}\Delta _E(x-x^{\prime })=\Delta
_E(0)=\frac{m^2}{(4\pi )^2}\left( \frac{m^2}{4\pi \mu ^2}\right) ^{\frac n2%
-2}\Gamma \left( 1-\frac n2\right)  \label{3.3}
\end{equation}
where $\mu $ is an arbitrary mass$.$ We can now define $\Delta _E^{(s)}(0)$,
the divergent component of $\Delta _E(0).$ As the $\Gamma \left( 1-\frac n2%
\right) $ behaves as: 
\begin{equation}
\Gamma \left( 1-\frac n2\right) \approx \frac 2{n-4}+\gamma  \label{3.4}
\end{equation}
when $n\rightarrow 4,$ (where $\gamma =\pi ^2/12$ is the Euler-Mascheroni
constant$),$ using the minimal substraction we find the singular part of $%
\Delta _E(0):$%
\begin{equation}
\Delta _E^{(s)}(0)=\frac{2m^2}{(4\pi )^2}\frac 1{n-4}  \label{3.5}
\end{equation}
In this way we have detected the local divergency. So we reach to a
decomposition (as (\ref{1.6'})-(\ref{1.7'})):

\begin{equation}
\Delta _E(0)=\Delta _E^{(r)}(0)+\Delta _E^{(s)}(0)=\frac{m^2}{(4\pi )^2}%
\left[ \left( \frac{m^2}{4\pi \mu ^2}\right) ^{\frac n2-2}\frac 12\Gamma
\left( 1-\frac n2\right) -\frac 2{n-4}\right] +\frac{2m^2}{(4\pi )^2}\frac 1{%
n-4}  \label{3.6}
\end{equation}

Then: 
\begin{equation}
\Delta _E^{(r)}(0)=\frac{m^2}{(4\pi )^2}\left[ \left( \frac{m^2}{4\pi \mu ^2}%
\right) ^{\frac n2-2}\frac 12\Gamma \left( 1-\frac n2\right) -\frac 2{n-4}%
\right]  \label{3.6'}
\end{equation}

Precisely when $n\rightarrow 4$ we have: 
\begin{equation}
\Delta _{E}^{(r)}(0)=\frac{m^{2}}{(4\pi )^{2}}\left[ \log \left( \frac{m^{2}%
}{4\pi \mu ^{2}}\right) +\gamma -1\right]  \label{3.7}
\end{equation}
where $\mu $ is the arbitrary mass, so essentially $\Delta _{E}^{(r)}(0)$
has an arbitrary value.

ii.-Let us now see how $\Delta _{E}^{(r)}(0)$ is related with mass
renormalization. In order to correct the divergency of $\langle \phi
_{0}(x)\phi _{0}(x^{\prime })\rangle $ we must correct the divergency of its
Fourier transform: 
\begin{equation}
(4.3.2)\qquad G_{0}(p)=\frac{1}{p^{2}+m_{0}^{2}+\Sigma _{0}(p)}  \label{3.7'}
\end{equation}

The computation of the tadpole graph (in the first $\lambda -$order) yields :

\begin{equation}
(4.3.7)\qquad \Sigma _{0}^{(1)}(p)=\frac{1}{2}\lambda _{0}\Delta _{E}(0)
\label{3.2}
\end{equation}
that makes the term $m_{0}^{2}+\Sigma _{0}^{(1)}(p)$ divergent. Precisely: 
\begin{equation}
m_{0}^{2}+\Sigma _{0}^{(1)}(p)=m_{0}^{2}+\frac{1}{2}\lambda _{0}\left[
\Delta _{E}^{(r)}(0)+\Delta _{E}^{(s)}(0)\right]  \label{3.2'}
\end{equation}
In usual renormalization we consider that the (bare) mass $m_{0}$ is
divergent. Then to compensate this divergency we define a (dressed) mass $m$
such that: 
\begin{equation}
m_{0}^{2}+\Sigma _{0}^{(1)}(p)=m^{2}+\Sigma ^{(1)}(p)  \label{3.2"}
\end{equation}
where both terms in the r.h.s. are finite, precisely: 
\begin{equation}
m_{0}^{2}=m^{2}\left[ 1-\frac{\lambda _{0}}{2}\Delta _{E}^{(s)}(0)\right]
\label{3.2'''}
\end{equation}
and 
\begin{equation}
\Sigma ^{(1)}(p)=\frac{1}{2}\lambda _{0}\Delta _{E}^{(r)}(0)  \label{3.7'x}
\end{equation}
Then the physical mass is: 
\begin{equation}
(4.3.15)\qquad m_{phys.}^{2}=m^{2}+\Sigma ^{(1)}(p)  \label{3.1}
\end{equation}
where $m_{phys}$ is a constant while $\Sigma ^{(1)}(p)$ and $m^{2},$ are
finite functions of $\mu $ (cf. eq. (\ref{3.7})) satisfying the
renormalization group equations.

iii.- Using the substraction recipe we would directly say that in eq. (\ref
{3.2'}) really $\Delta _{E}^{(s)}(0)=0$ and we will obtain: 
\begin{equation}
m_{phys.}^{2}=m_{0}^{2}+\frac{1}{2}\lambda _{0}\Delta _{E}^{(r)}(0)
\label{3.1'}
\end{equation}
which is equivalent to (\ref{3.1}) and where:

a.- $m_0$ plays the role of $m.$ It is therefore finite.

b.-Since $\Delta ^{(r)}(0)$ is a function of $\mu ,$ $m_{0}$ must also be a
function of $\mu $ in such a way that $m_{phys.}^{2}$ turns out to be a
constant. Then $m_{0}^{2}$ satisfies the same renormalization group equation
as the $m^{2}$ of eq. (\ref{3.1}).

This will be a common feature of substraction recipe for all physical
constants:{\it \ there is no need to introduce a dressed quantity since the
bare quantity takes its role, then the bare quantity becomes a function of }$%
\mu ${\it \ satisfying the renormalization group equations.}

\subsection{The cosmological constant and the Hadamard regularization for $%
\Delta _E(0)$ in the case $\lambda =0$.}

i.- Let us begin making an identification. $\Delta _{E}^{(r)}(0)$ in flat
space time can also be obtained in the case $n=4$ (but using Hadamard
substraction not minimal substraction) making all the curvatures zero in eq.
(\ref{2.19}), (namely making all the $"a"$ zero but $a_{0}=1$) and
multiplying by $-i$ (since $\Delta _{F}\rightarrow i\Delta _{E},$ \cite
{Brown} page. 194). So we obtain for $\lambda =0$:

\begin{equation}
\Delta _{E}^{(r)}(0)=\lim_{x,x^{\prime }\rightarrow 0}\Delta
_{E}^{(r)}(x,x^{\prime })=\frac{4lm^{2}}{(4\pi )^{2}}  \label{3.8}
\end{equation}
so essentially in this case $\lim_{x,x^{\prime }\rightarrow 0}\Delta
_{E}^{(r)}(x,x^{\prime })$ is just an arbitrary finite constant as in the
case of (\ref{3.7}). For the case $\lambda \neq 0$ some corrections will
appear in eq. (\ref{2.19}) (\cite{B&D} page 301) but the r.h.s. of eq. (\ref
{3.8}) will always be an arbitrary constant. The origin of this arbitrary is
the usual one: a infinite singularity can only be defined modulo a finite
undefined constant. So the arbitrary singularity coefficient $%
\lim_{x,x^{\prime }\rightarrow 0}\Delta _{E}^{(r)}(x,x^{\prime })$ defined
for $n=4$ plays the same role that $\mu $ in the case $n\neq 4.$ Both
parameters are related, when $\lambda =0,$ by: 
\begin{equation}
4l=\log \left( \frac{m^{2}}{4\pi \mu ^{2}}\right) +\gamma -1  \label{3.9}
\end{equation}
Thus, this preliminary consideration leads us to suppose that there must be
something like a cosmological constant also in $\lambda \phi ^{4}$ theory.
In fact: in traditional quantum field theory the additional infinite term
that appears, due to the addition of infinite ground energy terms $\frac{1}{2%
}\omega ,$ can be considered as an unrenormalizable cosmological constant.
This term is eliminated using normal ordering. But this renormalization is
better understood introducing the just mentioned cosmological constant (\cite
{Brown}, section 4.2) that must be renormalized. Using our equation we can
define a cosmological constant $\Lambda $ for this flat space-time theory,
if we just add to the usual lagrangian a term $\Lambda $ as we have done in
eq. (\ref{3.0})$.$ This term reads (see (\ref{2.20}) and (\ref{3.9}) in the
case $n=4$): 
\begin{equation}
\Lambda =\frac{m^{4}}{16\pi ^{2}}l=\frac{m^{4}}{4(4\pi )^{2}}\left[ \log
\left( \frac{m^{2}}{4\pi \mu ^{2}}\right) +\gamma -1\right]  \label{3.10}
\end{equation}

ii.- Let us see how renormalization method introduces the cosmological
constant. When $\lambda =0$ the vacuum to vacuum expectation (corresponding
to the vacuum one-loop graph) reads: 
\begin{equation}
(4.2.1)\qquad \langle 0+|0-\rangle =\int \left[ d\phi \right] \exp \left\{
-\int (d_{E}^{n}x)\left[ \frac{1}{2}(\partial _{\mu }\phi )^{2}+\frac{1}{2}%
m_{0}^{2}\phi ^{2}-\Lambda _{0}\right] \right\}  \label{3.12'}
\end{equation}
Thus: 
\begin{equation}
(4.2.2)\qquad \frac{\partial }{\partial m^{2}}\langle 0+|0-\rangle =-\frac{1%
}{2}\int (d_{E}^{n}x)\langle 0+|\phi (x)^{2}|0-\rangle =-\frac{1}{2}\langle
0+|0-\rangle \int (d_{E}^{n}x)\Delta _{E}(0)  \label{3.13'}
\end{equation}
and 
\begin{equation}
(4.2.4)\qquad \langle 0+|0-\rangle =\exp \left[ -\frac{1}{2}\int dm^{2}\int
(d_{E}^{n}x)\Delta _{E}(0)\right]  \label{3.10'}
\end{equation}
But if ${\cal E}$ is the cosmological energy density of the universe it also
is: 
\begin{equation}
(4.2.9)\qquad \langle 0+|0-\rangle =\exp \left[ -\int d_{E}^{n}x{\cal E}%
\right]  \label{3.10"}
\end{equation}
So: 
\begin{equation}
{\cal E=}\frac{1}{2}\int dm^{2}\Delta _{E}(0)-\Lambda _{0}  \label{3.14'}
\end{equation}
where $\Lambda _{0}$ can be considered as an integration constant. Then from
eq. (\ref{3.6}) we have: 
\begin{equation}
{\cal E=}\frac{1}{2}\int dm^{2}\Delta _{E}^{(r)}(0){\cal +}\frac{1}{2}\int
dm^{2}\Delta _{E}^{(s)}(0)-\Lambda _{0}  \label{3.15'}
\end{equation}
so we can consider that $\Lambda _{0}$ is infinite in such a way as to
cancel the infinite in $\Delta _{E}^{(s)}(0)$, namely: 
\begin{equation}
\Lambda _{0}=\frac{1}{2}\int dm^{2}\Delta _{E}^{(s)}(0)-\mu ^{4-n}\Lambda =%
\frac{1}{2}\frac{m_{0}^{2}}{(4\pi )^{2}}\frac{1}{n-4}-\mu ^{4-n}\Lambda
\label{3.16'}
\end{equation}
where $\Lambda $ is the finite cosmological constant. So finally: 
\begin{equation}
{\cal E=}\frac{1}{2}\int dm^{2}\Delta _{E}^{(r)}(0)+\mu ^{n-4}\Lambda
\label{3.17'}
\end{equation}
and when $n\rightarrow 4$ we have: 
\begin{equation}
(4.2.20)\qquad {\cal E=}\frac{1}{4}\frac{m^{4}}{(4\pi )^{2}}\left[ \ln
\left( \frac{m^{2}}{4\pi \mu ^{2}}\right) +\gamma -\frac{3}{2}\right]
-\Lambda  \label{3.A}
\end{equation}
where ${\cal E}$ is finite and it is not a function of $\mu $ but $\Lambda $
is a function of this mass. Using the Hadamard method of point i we can
directly see these facts using eq. (\ref{3.10}), since the $\mu $-variation
is cancelled in (\ref{3.A}). It remains a finite constant which is
unimportant since we can add an arbitrary constant to the lagrangian (\ref
{3.0}). As usual, the condition ${\cal E}=const.$ originates the
renormalization group equation for $\Lambda .$

iii.- Directly from (\ref{3.14'}) using substraction recipe we would have 
\begin{equation}
{\cal E=}\frac 12\int dm^2\Delta _E^{(r)}(0)+\Lambda _0  \label{3.14"}
\end{equation}
that for $n\rightarrow 4$ gives:

\begin{equation}
{\cal E=}\frac{1}{4}\frac{m^{4}}{(4\pi )^{2}}\left[ \log \left( \frac{m^{2}}{%
4\pi \mu ^{2}}\right) +\gamma -1\right] -\Lambda _{0}  \label{3.B}
\end{equation}
namely (\ref{3.A}) with the finite merely unimportant difference $\Delta 
{\cal E}=-\frac{m^{4}}{8(4\pi ^{2})},$ already discussed and $\Lambda _{0} $
playing the role of $\Lambda $. Now both terms in the r. h. s. are finite
and functions of $\mu $ while ${\cal E=E}_{phys}$ is a physical constant,
yielding the renormalization group equation for $\Lambda _{0}$ as in the
usual renormalization case.

From now on we will only use the dimensional regularization since the
singular structure of the higher point function is not so well studied as
the one of the two point function.

\section{First method. $\lambda \phi ^4$ theory at second perturbation order.
}

\subsection{Singular and regular parts of [$\Delta _F(z)]^{(d)2}$ and the
coupling constant renormalization.}

i.- Computing the fish graph we found that the scattering amplitude $T$
reads: 
\begin{equation}
(3.5.11)\qquad T=\lambda _{0}+\frac{1}{2}\lambda _{0}^{2}[F(s)+F(t)+F(u)]
\label{3.11}
\end{equation}
where $\lambda _{0}$ is the coupling constant and $s,t,u$ the Mandelstam
variables, $F$ reads \footnote{%
Really eq. (3.5.13) in ref. \cite{Brown} reads: 
\[
(3.5.13)\qquad F(-P^{2})=i\int (d^{4}z)e^{iPz}\Delta _{F}(z)^{2} 
\]
but we must remember that $\Delta _{F}(z)$ is a singular function (something
worse than a distribution) so $\Delta _{F}(z)^{2}$ is a meaningless
expression unless a multiplication procedure would be prescribed (which is
done in eq. (3.5.14) of ref. \cite{Brown}). Moreover decomposition (3.2.19),
of the same reference, which is the base of the equation above, cannot be
used when two points coincide, since this decomposition is inspired in the
case when these two points are far apart, as in the definition of the
truncated functions.}:

\begin{equation}
(3.5.13)\qquad F(-P^{2})=-\frac{i}{2}\int (d^{4}z)e^{iPz}\langle 0|T\phi
^{2}(0)\phi ^{2}(z)|0\rangle  \label{3.12}
\end{equation}
(as in ref. \cite{Brown} we have omitted the disconnected graphs, that were
taken into account in section III.B, and we will consider again in IV.B), $%
\langle 0|T\phi ^{2}(0)\phi ^{2}(z)|0\rangle $ is the four point function
divergent coincidence limit mentioned in section I.A that we must study and
substracted, precisely: 
\begin{equation}
(3.5.9)\qquad \langle 0|T\phi ^{2}(0)\phi ^{2}(z)|0\rangle =-2\Delta
_{F}(z)^{2}  \label{3.13}
\end{equation}
So we see that the coincidence limit is the (undefined) product of $\Delta
_{F}(z)$ by itself. Using dimensional regularization, as explained in
section I.A, we define 
\begin{equation}
\langle 0|T\phi ^{2}(0)\phi ^{2}(z)|0\rangle =-2\Delta _{F}(z)^{(d)2}
\label{3.13''}
\end{equation}
We will decompose this quantity as: 
\begin{equation}
\Delta _{F}(z)^{(d)2}=\Delta _{F}(z)^{(d)2(s)}+\Delta _{F}(z)^{(d)2(r)}
\label{3.14}
\end{equation}
according to the prescription (\ref{1.6})-(\ref{1.7}). Then we will obtain
the regular $F^{(r)}(-P^{2})$ as: 
\begin{equation}
F^{(r)}(-P^{2})=i\int (d^{4}z)e^{iPz}\Delta _{F}(z)^{(d)2(r)}  \label{3.15}
\end{equation}
and if we use this $F^{(r)}$ instead of $F$ in eq. (\ref{3.11}) the physical 
$T$ will turn out finite. We can directly make all the procedure on $%
F(-P^{2})$, the Fourier transform of $\Delta _{F}(z)^{2}.$ Using dimensional
regularization we obtain: 
\begin{equation}
(3.5.30)\qquad F(-P^{2})=-\frac{\mu ^{n-4}}{(4\pi )^{2}}\Gamma \left( 2-%
\frac{n}{2}\right) \int_{0}^{1}d\alpha \left[ \frac{m^{2}+\alpha (1-\alpha
)P^{2}}{4\pi \mu ^{2}}\right] ^{\frac{n}{2}-2}  \label{3.16}
\end{equation}
This equation can be considered as a way to obtain the square $\Delta
_{F}(z)^{2}$ i.e. to make this square when it is possible ($n\neq 4)$ and
then take the limit $n\rightarrow \infty .$ When $n\rightarrow 4$ it is: 
\begin{equation}
(3.5.31)\qquad \Gamma \left( 2-\frac{n}{2}\right) \rightarrow \frac{2}{4-n}%
+f(n)  \label{3.18}
\end{equation}
where $f(n)$ is a regular function such that $\lim_{n\rightarrow
4}f(n)=-\gamma .$ So, we can find the Fourier transform of the decomposition
(\ref{3.14}): 
\begin{equation}
F(-P^{2})=\mu ^{n-4}[F^{(s)}(-P^{2})+F^{(r)}(-P^{2})]  \label{3.19}
\end{equation}
where the factor $\mu ^{n-4}$ has been displayed to make $F^{(s)}(-P^{2})$
and $F^{(r)}(-P^{2})$ adimensional and where: 
\begin{equation}
F^{(s)}(-P^{2})=-\frac{1}{(4\pi )^{2}}\frac{2}{4-n}\int_{0}^{1}d\alpha =-%
\frac{1}{(4\pi )^{2}}\frac{2}{4-n}  \label{3.20}
\end{equation}
and 
\begin{equation}
(3.5.33)\qquad F^{(r)}(-P^{2})=-\frac{1}{(4\pi )^{2}}\int_{0}^{1}d\alpha
\left\{ \Gamma \left( 2-\frac{n}{2}\right) \left[ \frac{m^{2}+\alpha
(1-\alpha )P^{2}}{4\pi \mu ^{2}}\right] ^{\frac{n}{2}-2}+\frac{2}{n-4}%
\right\}  \label{3.21}
\end{equation}
Making now the inverse Fourier transforation of eq. (\ref{3.20}) we have
that: 
\begin{equation}
i\Delta _{F}(z)^{(d)2(s)}=\frac{1}{(4\pi )^{2}}\frac{2}{n-4}\delta (z)
\label{3.22}
\end{equation}
which, in fact, has the form announced in eq. (\ref{1.6}). It is singular
when $z=0$ and it shows that only the regular part is relevant when $z\neq
0. $ So we have detected the local divergency. Again, the non uniqueness of
the result is shown by the presence of $\mu $ in equation (\ref{3.19})$.$

ii.- The usual renormalization procedure would be to put the singular and
regular parts in (\ref{3.11}) to obtain: 
\begin{equation}
(3.5.37)\qquad T=\lambda _0+\lambda _0^2\frac{\mu ^{n-4}}{(4\pi )^2}\frac 3{%
n-4}+\frac{\mu ^{n-4}}2\lambda _0^2[F^{(r)}(s)+F^{(r)}(t)+F^{(r)}(u)]\}
\label{3.25}
\end{equation}
where the physical quantity $T$ must be finite and $\mu -$independent. This
is achieved introducing a renormalized $\lambda $ such that: 
\begin{equation}
(3.5.38)\qquad \lambda _0=\mu ^{4-n}\lambda \left( 1-\frac{3\lambda }{(4\pi
)^2}\frac 1{n-4}\right)  \label{3.26}
\end{equation}
so $\lambda _0$ turns out to be infinite and $\lambda $ finite: Then 
\begin{equation}
(3.5.48)\qquad T=\lambda +\frac 12\lambda
^2[F^{(r)}(s)+F^{(r)}(t)+F^{(r)}(u)]  \label{3.27}
\end{equation}
where all the magnitudes are finite. As $T$ is $\mu -$independent we can
obtain the renormalization group equation for $\lambda $.

iii.-According to the substraction method, we must make zero $\Delta
_{F}(z)^{(d)2(s)}$ or $F^{(s)}(-P^{2})$ and we obtain the finite physical
value of $T:$%
\begin{equation}
(3.5.48)\qquad T=\lambda _{0}+\frac{1}{2}\lambda
_{0}^{2}[F^{(r)}(s)+F^{(r)}(t)+F^{(r)}(u)]  \label{3.23}
\end{equation}
where $\lambda _{0}$ is a finite quantity.

Making the limit $n\rightarrow 4$ it turns out that: 
\begin{equation}
(3.5.66)\qquad F^{(r)}(s)=\frac 1{(4\pi )^2}\left\{ \log \left( \frac{%
m^2e^\gamma }{4\pi \mu ^2}\right) +\sqrt{1-\frac{4m^2}s}\log \left[ \frac{%
\sqrt{s-4m^2}-\sqrt{s}}{\sqrt{s-4m^2}+\sqrt{s}}-2\right] \right\}
\label{3.24}
\end{equation}

We see that with the substitution $\lambda _0\leftrightarrow \lambda $ eqs. (%
\ref{3.27}) and (\ref{3.23}) are the same. In the case of the substraction
method $\lambda _0$ is a finite $\mu -$function and as $T$ is $\mu -$%
independent so we can obtain the same renormalization group equation as
above.

\subsection{The cosmological corrected constant and $[\Delta _E(0)]^2$.}

i.- In the previous section we have neglected non-connected terms, e. g. in
eq. (\ref{3.2}), because the mass term was a consequence of the equation 
\begin{equation}
(4.3.5)\qquad G_0^{(1)}(x-x^{\prime })=-\frac 12\lambda _0\Delta _E(0)\int
(d_E^n\overline{x})\Delta _E(x-\overline{x})\Delta _E(x-\overline{x})
\label{4.1.1}
\end{equation}
that really reads: 
\begin{equation}
G_0^{(1)}(x-x^{\prime })=-\frac 12\lambda _0\int (d_E^n\overline{x})\left\{
\Delta _E(0)\Delta _E(x-\overline{x})\Delta _E(x-\overline{x})+\frac 1{4.3}%
\Delta _E(x-x^{\prime })\left[ \Delta _E(0)\right] ^2\right\}  \label{4.1.2}
\end{equation}

Moreover, the cosmological constant is originated in the equation:

\[
(3.3.9)\qquad \langle 0_{+}|0_{-}\rangle =\int \left[ d\phi \right] \exp
\left\{ -\int L_0(d_E^nx)\right\} \exp \left\{ -\frac{\lambda _0}{4!}\int
\phi ^4(d_E^nx)\right\} = 
\]
\[
\langle 0_{+}|0_{-}\rangle ^{(0)}-\frac{\lambda _0}{4!}\langle 0_{+}|\int
\phi ^4(d_E^nx)|0_{-}\rangle ^{(0)}+...= 
\]
\begin{equation}
\langle 0_{+}|0_{-}\rangle ^{(0)}-\frac{3\lambda _0}{4!}\int (d_E^n\overline{%
x})\left[ \Delta _E(0)\right] ^2\langle 0_{+}|0_{-}\rangle ^{(0)}
\label{4.1.3}
\end{equation}
which corresponds to eq. (\ref{3.10'}) with an extra term. $\langle
0_{+}|0_{-}\rangle ^{(0)}$ corresponds to the case $\lambda _0=0$ and the
second term to the non-connected graphs (the ''eight'', the ''square of the
figure eight'', etc.). In all these expressions $\left[ \Delta _E(0)\right]
^2$appears and it must be substituted by $\left[ \Delta _E(0)^{(r)}\right]
^2 $ according to the substraction recipe. As $\Delta _E(0)$ is not a
distribution, but just the divergent quantity (\ref{3.3}), we must only
substitute it by $\Delta _E^{(r)}(0)$, using decomposition (\ref{1.6'})-(\ref
{1.7'}), and making eqs. (\ref{4.1.2}) and (\ref{4.1.3}) finite.

ii.- Let us now go to the renormalization method: At order two we have: 
\[
(4.4.5)\qquad {\cal E}=\frac{m^n}{(4\pi )^{\frac n2}}\frac 1n\Gamma \left( 1-%
\frac n2\right) -\frac 12\mu ^{n-4}\frac{m^4}{(4\pi )^2}\frac 1{n-4}+ 
\]
\[
\frac 12\mu ^{n-4}\frac{\lambda m^4}{(4\pi )^2}\left[ \left( \frac{m^2}{4\pi
\mu ^2}\right) ^{\frac n2-2}\frac 12\Gamma \left( 1-\frac n2\right) -\frac 1{%
n-4}\right] ^2+ 
\]
\begin{equation}
\frac 12\mu ^{n-4}\frac{m^4}{(4\pi )^2}\frac 1{n-4}\left( 1-\frac \lambda {%
(4\pi )^2}\frac 1{n-4}\right) -\Lambda _0  \label{4.1.4}
\end{equation}
It can be checked that the two first lines of this equation are finite when $%
n\rightarrow 4.$ So we must define a renormalized cosmological constant $%
\Lambda $ such that: 
\begin{equation}
(4.4.6)\qquad \Lambda _0=\mu ^{n-4}\left[ \frac 12\frac{m^4}{(4\pi )^2}\frac %
1{n-4}\left( 1-\frac \lambda {(4\pi )^2}\frac 1{n-4}\right) +\Lambda \right]
\label{4.1.5}
\end{equation}
Then we have the final finite expression: 
\[
{\cal E}=\frac{m^n}{(4\pi )^{\frac n2}}\frac 1n\Gamma \left( 1-\frac n2%
\right) -\mu ^{n-4}\frac 12\frac{m^4}{(4\pi )^2}\frac 1{n-4}-\mu
^{n-4}\Lambda + 
\]
\begin{equation}
+\frac 12\mu ^{n-4}\frac{\lambda m^4}{(4\pi )^2}\left[ \left( \frac{m^2}{%
4\pi \mu ^2}\right) ^{\frac n2-2}\frac 12\Gamma \left( 1-\frac n2\right) -%
\frac 1{n-4}\right] ^2  \label{4.1.6}
\end{equation}
which is finite when $n\rightarrow 4$ . In fact, when $\lambda =0,$ we have
that:

\begin{equation}
{\cal E}=\frac{m^{n}}{(4\pi )^{\frac{n}{2}}}\frac{1}{n}\Gamma \left( 1-\frac{%
n}{2}\right) -\mu ^{n-4}\frac{1}{2}\frac{m_{0}^{4}}{(4\pi )^{2}}\frac{1}{n-4}%
-\mu ^{n-4}\Lambda  \label{4.1.7}
\end{equation}
which is a finite quantity, as we have proved in section III.B (it
corresponds to $\langle 0_{+}|0_{-}\rangle ^{(0)})$ while: 
\begin{equation}
\left[ \left( \frac{m_{0}^{2}}{4\pi \mu ^{2}}\right) ^{\frac{n}{2}-2}\frac{1%
}{2}\Gamma \left( 1-\frac{n}{2}\right) -\frac{1}{n-4}\right]  \label{4.1.8}
\end{equation}
is finite for (\ref{3.6'}), so the r. h. s. of eq. (\ref{4.1.6}) is finite.
When $n\rightarrow 4$ we find: 
\begin{equation}
{\cal E=}\frac{m^{4}}{4(4\pi )^{2}}\left[ \log \left( \frac{m^{2}}{4\pi \mu
^{2}}\right) +\gamma -\frac{3}{2}\right] +\frac{\lambda }{8}\frac{m^{4}}{%
(4\pi )^{2}}\left[ \log \left( \frac{m^{2}}{4\pi \mu ^{2}}\right) +\gamma
-1\right] ^{2}-\Lambda  \label{4.1.9}
\end{equation}
The terms of the r.h.s. are $\mu -$functions that originate the
renormalization group equation as usual.

iii.-Using directly the substraction method in eq. (\ref{4.1.4}) we would
have when $n\rightarrow 4:$%
\begin{equation}
{\cal E=}\frac{m^{4}}{4(4\pi )^{2}}\left[ \log \left( \frac{m^{2}}{4\pi \mu
^{2}}\right) +\gamma -1\right] +\frac{\lambda }{8}\frac{m^{4}}{(4\pi )^{2}}%
\left[ \log \left( \frac{m^{2}}{4\pi \mu ^{2}}\right) +\gamma -1\right]
^{2}-\Lambda _{0}  \label{4.1.10}
\end{equation}
with all terms finite and $\Lambda _{0}$ a function of $\mu $ as usual which
is equal to (\ref{4.1.9}) with the exception of the already known
unimportant constant. For both methods the renormalization group equation
can be obtained prescribing that ${\cal E}$ would not be a function of $\mu
. $

\subsection{The $\left[ \Delta _E(z)\right] ^{(d)3}$ and the wave function
renormalization.}

i.- Really mass renormalization of section III.A is based in the Green
function: 
\begin{equation}
(4.3.4)\qquad G_{0}^{(1)}(x-x^{\prime })=-\frac{1}{4!}\lambda _{0}\int
(d_{E}^{n}\overline{x})\langle \phi (x)\phi (x^{\prime })\phi ^{4}(\overline{%
x})\rangle ^{(0)}  \label{4.2.1}
\end{equation}
that can be written as: 
\begin{equation}
(4.3.5)\qquad G_{0}^{(1)}(x-x^{\prime })=-\frac{1}{4!}\lambda _{0}\Delta
_{E}(0)\int (d_{E}^{n}\overline{x})\Delta _{E}(x-\overline{x})\Delta
_{E}(x^{\prime }-\overline{x})  \label{4.2.2}
\end{equation}
In the next order we must compute: 
\begin{equation}
(4.5.2)\qquad G_{0}^{(2)}(x-x^{\prime })=\frac{1}{2}\left( -\frac{\lambda
_{0}}{4!}\right) \int (d_{E}^{n}y)(d_{E}^{n}z)\langle \phi (x)\phi
(x^{\prime })\phi ^{4}(y)\phi ^{4}(z)\rangle ^{(0)}  \label{4.2.3}
\end{equation}
Computing this Green function, as we have done with the previous one, we
find:

a.- Vacumm disconnected graphs: They are the ''eight'', the ''square
eight'', etc. which are removed by ordinary renormalization of the
cosmological constant or by the corresponding substraction\ that makes this
constant finite but undefined, as shown in eqs. (\ref{3.10}) or (\ref{4.1.10}%
), (\cite{Brown}, pages. 205 and 206).

b.- Disconnected two legs graph: It is the product of the ''tadpole'' by the
''eight''. Both graphs have already been considered either by
renormalization or substraction.

c.- Connected two legs graphs: Namely:

c$_1$.- The ''double scoop'' or ''double bubble'' graph, with an integral: 
\begin{equation}
(4.5.3)\qquad \Sigma _0^{(2,1)}(p)=-\frac 14\lambda _0^2\int (d_E^ny)\Delta
_E(y)^2\Delta _E(0)  \label{4.2.4}
\end{equation}
(which really is not a function of $p)$. It has two factors:

- $\Delta _E(0)$ that was considered in section III.A and made finite by
both recipes.

- $\Delta _E(y)^2=\Delta _E(y)^{(d)2}$ which was considered in section IV.A,
since the integral in eq. (\ref{4.2.4}) is just the integral in eq. (\ref
{3.15}) with $P=0$, which also was made finite by both recipes. So $\Sigma
_0^{(2,1)}$ turns out to be finite either way. Finally let us observe that
in $\Sigma _0^{(2,1)}$the typical expression $[w^{(2)}(0)]^\beta
[w^{(2)}(z)]^\alpha ,$ of section I.A, appears for the first time in its
complete version.

c$_2$.- The ''setting sun'' graph,which is a function of $p$%
\begin{equation}
(4.5.4)\qquad \Sigma _0^{(2,2)}(p)=-\frac 16\lambda _0^2\int (d_E^nx)\Delta
_E(x)^3e^{-ipx}  \label{4.2.5}
\end{equation}
To deal with this integral we must first compute $\Delta _E^{(d)3}(x)$
multiplying $\Delta _E(x)$ three times, then make its dimensional
regularization, and finally its Fourier transform $\Delta _E^{(d)3}(p).$ We
obtain: 
\begin{equation}
(4.5.37)\qquad \Sigma _0^{(2,2)}(p)=-\frac 16\left( \frac \lambda {(4\pi )^2}%
\right) ^2p^2\left( \frac{p^2}{4\pi \mu ^2}\right) ^{n-4}\frac{\Gamma \left( 
\frac n2-1\right) ^3\Gamma (3-n)}{\Gamma \left( \frac{3n}2-3\right) }
\label{4.3.3}
\end{equation}
As when $n\rightarrow 4$: 
\begin{equation}
\frac{\Gamma \left( \frac n2-1\right) ^3\Gamma (3-n)}{\Gamma \left( \frac{3n}%
2-3\right) }\rightarrow \frac 12\frac 1{n-4}  \label{4.3.4}
\end{equation}
then: 
\begin{equation}
\left[ \Sigma _0^{(2,2)}(p)\right] ^{(s)}=-\frac 1{12}\left( \frac \lambda {%
(4\pi )^2}\right) ^2p^2\frac 1{n-4}  \label{4.3.5}
\end{equation}

Then, we conclude that: 
\begin{equation}
\Delta _E^{(d)3(s)}(x)=\frac 12\left( \frac 1{2\pi }\right) ^n\frac 1{(4\pi
)^2}\frac 1{n-4}\int p^2e^{ipx}(d_E^np)=\frac 12\frac 1{(4\pi )^2}\frac 1{n-4%
}\nabla ^2\delta (x)  \label{4.2.7}
\end{equation}
which, in fact, has the form announced in eq. (\ref{1.6}). It is local,
since it is singular when $z=0$ and vanishing for $z\neq 0.$ We have
detected another local singularity. In the finite limit $n\rightarrow 4$ we
obtain

\begin{equation}
(4.5.38)\qquad \Sigma _0^{(2,2)}(p)=-\frac 1{12}\left[ \frac \lambda {(4\pi
)^2}\right] ^2p^2\left( \log \frac{p^2}{4\pi \mu ^2}+const.\right)
\label{4.2.6}
\end{equation}
Substracting all singularities the propagator $G_0^{(2)}(x-x^{\prime })$
turns out to be finite to the second $\lambda -$order. But, of course, an
ambiguity appears in the constant of eq. (\ref{4.2.6}) that must be fixed by
a measurement.

ii.- In the renormalization theory we must add all the results of the
connected graphs to obtain: 
\[
(4.5.39)\qquad G_{0}(p)^{-1}=p^{2}\left\{ 1-\frac{1}{12}\left[ \frac{\lambda 
}{(4\pi )^{2}}\right] ^{2}\frac{1}{n-4}\right\} +m_{0}^{2}+m^{2}\left\{ 
\frac{\lambda }{(4\pi )^{2}}\left[ \frac{1}{n-4}+finite\right] \right\} - 
\]
\begin{equation}
-\left[ \frac{\lambda }{(4\pi )^{2}}\right] ^{2}\left\{ m^{2}\left[ \frac{2}{%
(n-4)^{2}}+\frac{1}{2}\frac{1}{n-4}\right] +finite\text{ }function\text{ }of%
\text{ }p^{2}\right\}  \label{4.3.7}
\end{equation}

To eliminate the infinities via renormalization a renormalized $G(p)$ is
defined as: 
\begin{equation}
(4.5.40)\qquad G_{0}(p)=z_{1}^{2}G(p)  \label{4.3.8}
\end{equation}
where: 
\begin{equation}
(4.5.41)\qquad z_{1}^{2}=1+\frac{1}{12}\left[ \frac{\lambda }{(4\pi )^{2}}%
\right] ^{2}\frac{1}{n-4}  \label{4.3.9}
\end{equation}
then up to the order $\lambda ^{2}$ we have: 
\[
(4.5.42)\qquad G(p)^{-1}=p^{2}+m_{0}^{2}+m^{2}\left\{ \frac{\lambda }{(4\pi
)^{2}}\left[ \frac{1}{n-4}+finite\right] \right\} - 
\]
\begin{equation}
-\left[ \frac{\lambda }{(4\pi )^{2}}\right] ^{2}\left\{ m^{2}\left[ \frac{2}{%
(n-4)^{2}}+\frac{5}{12}\frac{1}{n-4}\right] +finite\text{ }\right\}
\label{4.3.10}
\end{equation}
and the renormalized the mass reads: 
\begin{equation}
(4.5.43)\qquad m_{0}^{2}=m^{2}\left\{ 1-\frac{\lambda }{(4\pi )^{2}}\frac{1}{%
n-4}+\left[ \frac{\lambda }{(4\pi )^{2}}\right] ^{2}\left[ \frac{2}{(n-4)^{2}%
}+\frac{5}{12}\frac{1}{n-4}\right] \right\}  \label{4.3.11}
\end{equation}
Then: 
\begin{equation}
G(p)^{-1}=p^{2}+m^{2}+finite(\mu )  \label{4.3.12}
\end{equation}
All terms are finite and $G(p)^{-1}$, $m^{2},$ and $finite(\mu )$are $\mu -$%
functions. $p^{2}$ is the physical constant quantity that originates the
renormalization group. The renormalization of eqs. (\ref{4.3.8}) and (\ref
{4.3.9}) is usually considered as a wave function renormalization: 
\begin{equation}
(4.5.46)\qquad \phi _{0}(x)=z_{1}\phi (x)  \label{4.3.13}
\end{equation}
where $\phi (x)$ is the renormalized field.

iii.- Using the substraction recipe eq. (\ref{4.3.12}) reads: 
\begin{equation}
G(p)^{-1}=p^{2}+m_{0}^{2}+finite(\mu )  \label{4.3.14}
\end{equation}
since all the infinities disappear from eqs. (\ref{4.3.10}) and (\ref{4.3.11}%
), but a finite undeterminate constant remains that must be fixed by a
measurement that corresponds to the one of the wave function
renormalization. As usual $G(p)^{-1}$, $m_{0}^{2}$ and $finite(\mu )$ are $%
\mu -$functions that originate the renormalization group. Moreover, from eq.
(\ref{4.3.9}) with no infinity we have $z_{1}=1$ and there is no need of the
wave function renormalization\footnote{%
This fact must be most welcome since now both the ''bare'' and
''renormalized'' fields satisfy the same equal time commutation relations.}.
Then using our recipe, the result is the same.

\section{First method. $\lambda \phi ^4$ theory at any order. More general $%
\lambda \phi ^l$ theories. Speculations on non-renormalizable theories}

We can now follow a well known path. For the $\lambda \phi ^{4}$ theory the
superficial divergence is: 
\begin{equation}
(5.2.21)\qquad D=4-N  \label{5.1}
\end{equation}
where $N$ is the number of external legs of the graph. Then, only graphs
with $N=2$ and $N=4$ have basic divergencies. Moreover the convergence of
all the graphs, to $\lambda ^{2}$ can be reduced to prove the convergence of
the primitive divergent graph (\cite{Ramond}, page. 144), namely the tadpole
and the fish graphs, the double scoop, and, the setting sun (and the non
connected graphs) which were studied in the previous sections. These graphs
are finite under ordinary renormalization (or if the substraction recipe are
used). So, repeating these calculations to any order all graphs of the
renormalized theory are finite and the theory turns out to be finite to all
orders \cite{Renor}. $\lambda \phi ^{4}$ theory can be considered as
renormalizable since it has a finite number of primitive divergent graphs
and therefore a finite number of relevant singular point functions, namely
three. So we now know that{\it \ using substraction method the theory is
directly finite to any order}.

To complete the panorama we can study the problem in more general scalar
field theories. Theories with interactions $\lambda \phi ^{l}$ with $l>4$
turn out to be non renormalizable because they have an infinite number of
primitive divergent graphs and therefore a infinite number of relevant
singular point functions, that cannot be compensated with the finite number
of terms of the bare lagrangian. But the substraction recipe can anyhow be
used making all these singular functions finite, and these theories would
become also finite. So {\it all theories can be made finite if we use the
substraction recipe.}

In fact, let us consider what we know about this kind of theories:

i.-In order to make the theory finite we must make finite (by
renormalization or substraction) all the superficially divergent subgraphs ($%
D\geq 0).$ The mass dimension in each term is the superficial divergency.

ii.- The divergent terms are polynomials of finite order in the external
momentum. Using dimensional regularization with minimal substraction the
coefficients of these polynomials contain positive integer powers of the
parameters of the theory multiplied by poles in $n-4$ (\cite{Brown}, page
235). So the typical {\it divergent} term reads: 
\begin{equation}
P(p_{1},p_{2},...,p_{N})=\sum A_{\alpha \beta \delta _{1},...\delta
_{N}}^{\gamma }\frac{m^{\alpha }\lambda ^{\beta }...}{(n-4)^{\gamma }}%
p_{1}^{\delta _{1}}p_{2}^{\delta _{2}}...p_{N}^{\delta _{N}}  \label{5.2}
\end{equation}
that, under a Fourier transform: $w_{N}^{(s)}(x_{1},x_{2},...,x_{N})\approx $

$\int dp_{1}\int dp_{2}...\int
dp_{N}P(p_{1},p_{2},...,p_{N})e^{-ix_{1}p_{1}}e^{-ix_{2}p_{2}}...e^{-ix_{N}p_{N}} 
$ corresponds to the local singularity: 
\begin{equation}
w_{N}^{(s)}(x_{1},x_{2},...,x_{N})\approx \sum A_{\alpha \beta \delta
_{1},...\delta _{N}}^{\gamma }\frac{m^{\alpha }\lambda ^{\beta }...}{%
(n-4)^{\gamma }}\nabla ^{\delta _{1}}\delta (x_{1})\nabla ^{\delta
_{2}}\delta (x_{2})...\nabla ^{\delta _{N}}\delta (x_{N})  \label{5.3}
\end{equation}
as in eq. (\ref{4.2.7}), i. e. singularities of the (\ref{1.6'}) type. All
the singularities $\nabla ^{\delta i}\delta (x_{i})$ are well defined
distributions in variable $x_{i}$ (there are no meaningless expressions as $%
\delta (0)\int_{0}^{\infty }d\omega $ that we will consider and eliminate in
the next section) multiplied by infinite poles $1/(n-4)^{\gamma }$.

So let us compare the two methods:

i.- Renormalization: In this case the divergent (\ref{5.2}) terms must be
compensated by counterterms like: 
\begin{equation}
\frac{\delta m^{\alpha ^{\prime }}\delta \lambda ^{\beta ^{\prime }}...}{%
(n-4)^{\gamma }}p_{1}^{\delta _{1}}p_{2}^{\delta _{2}}...p_{N}^{\delta _{N}}
\label{5.4}
\end{equation}

where $\alpha \neq \alpha ^{\prime },\beta \neq \beta ^{\prime }$ but $%
\alpha +\beta =\alpha ^{\prime }+\beta ^{\prime }$ in such a way to have the
same dimension (or the same superficial divergence $D).$ It is clear that in
general such counter terms must be infinite and will be only finite in
particular cases (renormalizable theories). Moreover non-renormalizable
theories are considered non-controllable, since they must have an infinite
number of counter terms, implying new interaction terms of growing power.

ii.- Divergencies will disappear using the substraction recipe and the
theory will turn out finite anyhow. In fact, as in our method the lagrangian
remains untouched, and we can make the theory finite simply substracting the
divergent terms. Then all the $\gamma >0$ terms will disappear and $%
w_{N}^{(r)}(x_{1},x_{2},...,x_{N})$ will be a well defined functions 
\footnote{%
Really we also have an infinite set of counter terms, but not in the
lagrangian, they are the singular terms of the point functions that must we
substracted from these functions to obtain the regular terms, so they are
precisely located in the place where they are needed.}. As, from the general
formalism of quantum theory \cite{Haag}, we are used to deal with a host of
infinite divergent point functions\footnote{%
Like those listed in footnote 1.}, to deal with a similar host of finite
point functions, obtained via the substraction recipe, it cannot be a major
theoretical problem. So under our method both renormalizable and non-
renormalizable theories are finite. Nevertheless, in renormalizable theories
the ambiguous terms are combined in such a way that the unknown parameters
of the theory can be computed with a finite number of physical data, while
in the case of non-renormalizable theories this number is infinite\footnote{%
E. g.: in the $\lambda \phi ^{4}$ theory the renormalization group shows
that all the residues of the poles depend on those of the first order poles (%
\cite{Brown}. page 241). Namely all the ambiguities corresponding to higher
divergencies depend on the first order ambiguities, and therefore all these
ambiguities can be computed with just some measurements. In the general $%
\lambda \phi ^{l}$ case there is not such a miracle and we must deal with
infinite ambiguities.}.

Then using our method, non-renormalizable theories most likely make some
sense and, if they have small coupling constants, probably would yield good
results, using a few terms of the perturbation expansion and a few physical
data, but of course we do not know yet if they have any physical relevance.
Moreover, in recent years it has become increasingly apparent that the usual
renormalization is not a fundamental physical requirement (\cite{Weinberg},
vol. 1, page 518). We stop our speculation here, since this will be the
subject of forthcoming researches.

\section{Second method.}

In this section we will try to find a theoretical justification for the
substraction method, following the authors' ideas of the references: \cite
{Haag}, \cite{Segal}, \cite{Bogo}, and, \cite{VanHowe}. We will also find
new potentially dangerous divergencies hidden in the formalism that will
also be eliminated. The quoted authors consider that the first object that
must be taken into account in quantum fields theory is the set of
observables $O$ that we will use (belonging to the space of the relevant
observables ${\cal O}$). Then the states $\rho $ can be considered as the
functionals over these observables yielding the mean values $(\rho |O).$ If
the spectra of the observables of the problem are discrete we have that $%
(\rho |O)=Tr(\rho O).$ If one or many of these spectra are continuous the
problem is more difficult because the last symbol is ill-defined. This
happens, e. g., when the energy spectrum is continuous. In papers \cite
{Laura} we solve this problem (based in the mathematical structure
introduced in paper \cite{ALTS}), finding good results for many physical
problems. In the present paper we deal with short distance divergences,
related with the position operators, which also have a continuous spectrum.
So we will try to adapt the method of paper \cite{Laura} to this new
problem. But first let us review the formalins of this paper.

\subsection{Van Hove formalism.}

Let us consider a system with an hamiltonian $H$ with continuous energy
spectrum $0\leq \omega <+\infty .$ In the simple case at least some
generalized observables reads: 
\begin{equation}
O=\int \int d\omega d\omega ^{\prime }\left[ O_{\omega }\delta (\omega
-\omega ^{\prime })+O_{\omega \omega ^{\prime }}\right] |\omega \rangle
\langle \omega ^{\prime }|  \label{6.1}
\end{equation}
where $O_{\omega }$ and $O_{\omega \omega ^{\prime }}$ are regular functions
(with properties we will discuss below). These observables are contained in
a space ${\cal O.}$ The introduction of distributions like $\delta (\omega
-\omega ^{\prime })$ is necessary because the ''singular term'' $O_{\omega
}\delta (\omega -\omega ^{\prime })$ appears in observables that cannot be
left outside the space ${\cal O}$, like the identity operator, the
hamiltonian operator, or the operators that commute with the hamiltonian.
So, even in this simple case the observables contain $\delta $ functions
(while in more elaborated cases they will also contain other kind of
distributions). Symmetrically a generalized state reads: 
\begin{equation}
\rho =\int \int d\omega d\omega ^{\prime }\left[ \rho _{\omega }\delta
(\omega -\omega ^{\prime })+\rho _{\omega \omega ^{\prime }}\right] |\omega
\rangle \langle \omega ^{\prime }|  \label{6.2}
\end{equation}
where $\rho _{\omega }$ and $\rho _{\omega \omega ^{\prime }}$ are also
regular functions (with properties to be defined). These states are
contained in a convex set of states ${\cal S.}$ The introduction of
distributions like $\delta (\omega -\omega ^{\prime })$ is also necessary in
this case because the ''singular term'' $\rho _{\omega }\delta (\omega
-\omega ^{\prime })$ appears in generalized states that cannot be left
outside the set ${\cal S}$, like the equilibrium state\footnote{%
Usually this state is not considered in the scattering theory. So it is only 
{\it potentially dangerous} for more general theories}. With this
mathematical structure it is impossible to calculate something like $Tr(\rho
O)$ because meaningless $\delta (0)\int_{0}^{\infty }d\omega $ appear. {\it %
This is the main problem (if }$O_{\omega }\neq 0$ {\it and }$\rho _{\omega
}\neq 0)$. Let us keep in mind that with the old philosophy we are just
considering the mean value $Tr(\rho O)$ as a simple {\it inner product} (and
in doing so we have the problem of the $\delta (0)\int_{0}^{\infty }d\omega
) ${\it .}

The problem is solved if we consider the characteristic algebra of the
operators ${\cal A}$ (see the complete version in \cite{Ordo}) containing
the space of the self-adjoints observables ${\cal A}_{S}$ which contains the
minimal subalgebra $\widehat{{\cal A}\text{ }}$of the observables that
commute with the hamiltonian $H$ (that we can consider as the typical
''diagonal'' operators). Then we have: 
\begin{equation}
\widehat{{\cal A}\text{ }}\subset {\cal A}_{S}{\cal \subset A}  \label{A1}
\end{equation}
Now we can make the quotient 
\begin{equation}
\frac{{\cal A}}{\widehat{{\cal A}}}={\cal V}_{nd}  \label{A1'}
\end{equation}
where ${\cal V}_{nd}$ would represent the vector space of equivalent classes
of operators that do not commute with $H$ (the ''non-diagonal operators'').
These equivalence classes reads 
\begin{equation}
\lbrack a]=a+\widehat{{\cal A}},\qquad a\in {\cal A,}\text{ }[a,H]\neq 0
\label{A2}
\end{equation}
So we can decompose ${\cal A}$ as: 
\begin{equation}
{\cal A=}\widehat{{\cal A}}+{\cal V}_{nd}  \label{A3}
\end{equation}
(this decomposition corresponds to the one in eq. (\ref{6.1}). But neither
the two + of the last two equation is a direct sum, since we can add and
substract an arbitrary $a\in \widehat{{\cal A}}$ from each term of the r. h.
s. of the last equation.

At this point we can ask ourselves which are the measurement apparatuses
that really matter in the case of decoherence under a evolution $e^{-iHt}$.
Certainly the apparatuses that measure the observables that commute with $H$
which are contained in $\widehat{{\cal A}}$ that correspond to diagonal
matrices $\sim \delta (\omega -\omega ^{\prime }).$ Also the apparatuses
that measure observables that do not commute with $H$ that corresponds to
the off-diagonal terms that are contained in ${\cal V}_{nd}.$ The terms
corresponding to the latter kind of apparatuses (either in the observables
or in the corresponding states) must vanish when $t\rightarrow \infty ,$ so
they must be endowed with mathematical properties adequated to produce this
limit. Riemann-Lebesgue theorem tell us that this fact take places if
functions $O_{\omega \omega ^{\prime }}$ are regular (and also the $\rho
_{\omega \omega \acute{}}$ below). So we define a sub algebra of ${\cal A} $%
, that can be called a van Hove algebra, as:

\begin{equation}
{\cal A}_{vh}{\cal =}\widehat{{\cal A}}\oplus {\cal V}_{r}\subset {\cal A}
\label{A4}
\end{equation}
where the vector space ${\cal V}_{r}$ is the space of observables with $%
O_{\omega }=0$ and $O_{\omega \omega ^{\prime }}$ a regular function. Now
the $\oplus $ is a direct sum because $\widehat{{\cal A}}$ contains $\delta
(\omega -\omega ^{\prime })$ and ${\cal V}_{r}$ just regular functions and a
kernel cannot be both a $\delta $ and a regular function. Moreover, as our
observables must be selfadjoint the space of observables must be 
\begin{equation}
{\cal O=A}_{vhS}{\cal =}\widehat{{\cal A}}\oplus {\cal V}_{rS}\subset {\cal A%
}_{S}  \label{A5}
\end{equation}
where ${\cal V}_{rS}$ contains only self-adjoint operator (namely $O_{\omega
\omega ^{\prime }}^{*}$=$O_{\omega ^{\prime }\omega }).$ Restriction (\ref
{A5}) is just the choice (coarse-graining) of the relevant measurement
apparatuses for our problem, those that measure the diagonal terms in $%
\widehat{{\cal A}}$ and those that measure the non diagonal terms that
vanish when $t\rightarrow \infty $ in ${\cal V}_{rS}.$ Moreover ${\cal O=A}%
_{vhS}$ is dense in ${\cal A}_{S}$ (because any distribution can be
approximated by regular functions) and therefore essentially it is the
minimal possible coarse-graining. Let us call $|\omega )=|\omega \rangle
\langle \omega |$ to the vectors of the basis of $\widehat{{\cal A}}$ and $%
|\omega ,\omega ^{\prime })=|\omega \rangle \langle \omega ^{\prime }|$ to
those of ${\cal V}_{rS}$. Then a generic observable of ${\cal O}$ reads 
\begin{equation}
O=\int d\omega O_{\omega }|\omega )+\int \int d\omega d\omega ^{\prime
}O_{\omega ^{\prime }}|\omega ,\omega ^{\prime })  \label{6.3}
\end{equation}
namely is a vector in the basis $\{|\omega ),|\omega ,\omega ^{\prime })\}$
where $O_{\omega }$ and $O_{\omega \omega ^{\prime }}$ are regular functions
(with properties precised in the paper \cite{Laura} and omitted here, as we
will do with all the functions that will appear in this brief review).{\it \ 
}

The states must be considered as linear functional over the space ${\cal O}$
(${\cal O}^{\prime }$ the dual of space ${\cal O}$\cite{Segal}, \cite{Bogo}, 
\cite{VanHowe}):

\begin{equation}
{\cal O}^{\prime }{\cal =A}_{vhS}^{\prime }{\cal =}\widehat{{\cal A}}%
^{\prime }\oplus {\cal V}_{rS}^{\prime }\subset {\cal A}_{S}^{\prime }
\label{A6}
\end{equation}
Therefore the states read: 
\begin{equation}
\rho =\int d\omega \rho _{\omega }(\omega |+\int \int d\omega d\omega
^{\prime }\rho _{\omega \omega ^{\prime }}(\omega ,\omega ^{\prime }|
\label{6.4}
\end{equation}
where $\rho _{\omega }$ and $\rho _{\omega \omega ^{\prime }}$ are regular
functions and $\{(\omega |,(\omega ,\omega ^{\prime }|\}$ is the cobasis of $%
\{|\omega ),|\omega ,\omega ^{\prime })\}$. The set of these generalized
states is the convex set ${\cal S\subset O}^{\prime }.$ Now the mean value: 
\begin{equation}
(\rho |O)=\int d\omega \rho _{\omega }O_{\omega }+\int \int d\omega d\omega
^{\prime }\rho _{\omega \omega ^{\prime }}O_{\omega ^{\prime }\omega }
\label{6.5}
\end{equation}
is well defined and yields reasonable physical results \cite{Laura}\footnote{%
Moreover, the introduction of the singular observables automatically yield
the introduction of the singular states \cite{Laura}.}. In the last equation
terms like $\delta (0)\int_{0}^{\infty }d\omega $ have disappeared. {\it %
This is the simple trick that allows as to deal with the singularities in a
rigorous mathematical way and to obtain correct physical results in papers }%
\cite{Laura} {\it and} \cite{Deco}. Essentially we have defined a new
observable space ${\cal O}$ that contains the observables $O$ of eq. (\ref
{6.3}) that are adapted to solve our problem. In this way we have found a
method to deal with the singular terms containing Dirac's deltas. We are now
considering the mean value $(\rho |O)$ not as an inner product but as a the
action {\it functional} $\rho $ acting on the vector $O$ (and the $\delta
(0)\int_{0}^{\infty }d\omega $ have disappeared)$.$ Decoherence is a
consequence of Riemann-Lebesgue theorem in the time evolution of the last
equation, namely: 
\begin{equation}
(\rho (t)|O)=\int d\omega \rho _{\omega }O_{\omega }+\int \int d\omega
d\omega ^{\prime }e^{-i(\omega -\omega ^{\prime })t}\rho _{\omega \omega
^{\prime }}O_{\omega ^{\prime }\omega }
\end{equation}

\subsection{The formalism in the simplest case.}

Let us now use the same technique to deal with the singularities of quantum
field theory. But first let us remember that in quantum field theory there
coexist at least two different mathematical structures:

- The abstract Hilbert space ${\cal H}$ where the field $\phi (x)$ is an
operator and the vacuum state $|0\rangle $ a vector. The multiplication in
the characteristic algebra ${\cal A}$ is the multiplication of these
operators. This is not the place where divergencies are produced. Therefore
we will not modify this structure.

- The vector space of functions of $N$ ($N\rightarrow \infty )$ variables $%
x_{1},x_{2},...,x_{N}$ where the functions $\phi (x_{1})\phi (x_{2})...\phi
(x_{N})$ can be considered as the coordinates of the vectors of a vector
space ${\cal N}$ in a basis $|x_{1},x_{2},...,x_{N}).$ Since we have proved
that really the ''functions'' $\phi (x_{1})\phi (x_{2})...\phi (x_{N}) $ are
distributions or worse we will give to this space the mathematical structure
that we explained in the previous subsection\footnote{%
Mathematically speaking this would be the one of a ''nuclear'' space ${\cal N%
}$ , namely the generalization of the ordinary N-rank tensor space to the
case where the $N$ indices are continuous. In the future we will base an
axiomatic quantum field theory using this mathematical structure.}.

The characteristic algebra is ${\cal A=H\otimes H\otimes N.}$

Let us begin with the case of just two variables to see the analogy with the
previous section. Then, as the observables like $\phi (x)\phi (x^{\prime })$
are distributions (or worse) it is reasonable to consider that all the
observables are singular\footnote{%
We may say that we are using the continuous spectrum of the position
operator \cite{Gaioli} which is $-\infty <x<+\infty $ and define the basis $%
|x,x^{\prime })$ as $|x\rangle \langle x|.$ But this is not necessary since
we can directly say that the space ${\cal N}$ of vectors with coordinates $%
\phi (x)\phi (x^{\prime })$ has a basis $|x,x^{\prime }).$}. Let us begin
with the simplest case i. e.: with just the singularity (\ref{3.22}). Then
our observables would read (like in (\ref{6.1}) or (\ref{3.22}) with $%
z=x-x^{\prime }$): 
\begin{equation}
O=\int \int dxdx^{\prime }\left[ \frac{O_{x}}{n-4}\delta (x-x^{\prime
})+O_{xx^{\prime }}\right] |x,x^{\prime })  \label{6.6}
\end{equation}
where $O_{x}$ and $O_{xx^{\prime }}$ are regular functions. But if we
continue the road of eqs. (\ref{6.1}) and (\ref{6.2}) we will find the same
problems as above. On the other hand using the philosophy just explained%
\footnote{%
But now referred to the measurement apparatuses, i. e. those that measure
variable $x$ that now take the role of variable $\omega .$} we can define
the space of observables

\begin{equation}
{\cal O=A}_{vhS}{\cal =}\widehat{{\cal A}}\oplus {\cal V}_{rS}\subset {\cal A%
}_{S}  \label{A7}
\end{equation}
where $\widehat{{\cal A}}$ is now the space of the $\delta (x-x\acute{})-$%
singularity with pole $(n-4)^{-1}$ and ${\cal V}_{nS\text{ }}$ is the space
of regular observables measured by physical apparatuses. $O_{x}$ and $%
O_{xx^{\prime }}$ are regular function and ${\cal O=A}_{vhS}$ is dense in $%
{\cal A}_{S}.$ Then we may transform the eq. (\ref{6.6}) to make it similar
to (\ref{6.3}), namely: 
\begin{equation}
O=\int dx\frac{O_{x}}{n-4}|x)+\int \int dxdx^{\prime }O_{xx^{\prime
}}|x,x^{\prime })  \label{6.7}
\end{equation}
so now the observables are vectors of a space ${\cal O\subset N\otimes
H\otimes H}$ of basis $\{|x),|x,x^{\prime })\}.$ Then the states of this
system are just some linear functional over the space ${\cal O.}$

\begin{equation}
{\cal O}^{\prime }{\cal =A}_{vhS}^{\prime }{\cal =}\widehat{{\cal A}}%
^{\prime }\oplus {\cal V}_{rS}^{\prime }\subset {\cal A}_S^{\prime }
\label{A8}
\end{equation}
For a moment let us postulate that also the singularities in the states do
exist \footnote{%
This is not really the case as we will see in the next subsection.}. In this
perspective the state must be linear combinations in the basis $%
\{(x|,(x,x^{\prime }|\}$ $($where $\{(x|,(x,x^{\prime }|\}$ is the cobasis
of $\{|x),|x,x^{\prime })\})$, so they must read:

\begin{equation}
\rho =\int dx\rho _{x}(x|+\int \int dxdx^{\prime }\rho _{xx^{\prime
}}(x,x^{\prime }|  \label{6.8}
\end{equation}
where $\rho _{x}$ and $\rho _{xx^{\prime }}$ are regular function. With
these definitions the action of functional $(\rho |$ over the vector $|O)$
reads: 
\begin{equation}
(\rho |O)=\int dx\frac{\rho _{x}O_{x}}{n-4}+\int \int dxdx^{\prime }\rho
_{xx^{\prime }}O_{x^{\prime }x}  \label{6.8'}
\end{equation}
and it will be well defined when $n=4$ only if the first term of the r.h.s.
vanishes. But this is precisely the case since, based in the arguments of
subsection I.B, we know that either the real physical observables must be
such that $O_{x}=0,$ namely they cannot see the singularities of the states
(because really they only are mathematical artifacts, etc.) or $\rho _{x}=0$
(namely the states cannot see the singularities of the observables, etc.).
Then either $O_{x}=0$ or $\rho _{x}=0$ and the last equation reads: 
\begin{equation}
(\rho |O)=\int \int dxdx^{\prime }\rho _{xx^{\prime }}O_{x^{\prime }x}
\label{6.10}
\end{equation}
and therefore we have eliminated the singular term $\frac{\rho _{x}O_{x}}{n-4%
}$ of eq. (\ref{6.8'}) which now have no physical effect.{\it \ In this way
we can justified the elimination of {\bf all} singular terms as we have done
with} (\ref{3.22}) {\it as we will see }\footnote{%
Of course we can also directly say that the term $\int dx\rho
_{x}O_{x}/(n-4) $ is unphysical. But there is a difference between eq. (\ref
{6.5}) and the last equation. In the former the singular observables see the
singular states and therefore it has two terms. In the latter there are
either singular observables or singular states and it has only one term.
Therefore the two coarse-graining use in sections VI. A and VI.B are
different. This fact is no surprising since the singular terms (in $\omega )$
are necessary in the case of decoherence to represent the diagonal final
state but these singular terms (in $x)$ must disappear in the case of
quantum field theory since this is the way divergent poles disappear. The
two different coarse-graining are introduced to explain two different
observed physical facts.}.

\subsection{The formalism in the general case.}

To generalize this idea let us go back to eq. (\ref{1.3}). We know that the
functional $Z[\rho ]$ and its derivatives define the whole theory. Moreover,
following the above ideas it must be written as\footnote{%
The next symbol contains a sum over the indices $N=0,1,2,...$}: 
\begin{equation}
Z[\rho ]=\exp i(\rho |O)  \label{6.10'}
\end{equation}
where 
\begin{equation}
|O)=|\phi (x_{1})\phi (x_{2})...\phi (x_{N}))  \label{6.10''}
\end{equation}
being $\phi (x_{1})\phi (x_{2})...\phi (x_{N})$ the components of a vector $%
|O)\in {\cal A=N\otimes H\otimes H}$ for any $N$ and 
\begin{equation}
(\rho |=\rho (x_{1})\rho (x_{2})...\rho (x_{N})|0\rangle \langle 0|
\label{6.10'''}
\end{equation}
where $(\rho |\in {\cal A}^{\prime }{\cal =N}^{\prime }{\cal \otimes
H\otimes H.}$ Remember that what really matters for our analysis are the
''functions'' $\phi (x_{1})\phi (x_{2})...\phi (x_{N})$ and $\rho
(x_{1})\rho (x_{2})...\rho (x_{N})$ are in spaces ${\cal N}$ and ${\cal N}%
^{\prime }$ while the way to operate with $|0\rangle \langle 0|$ over the
field $\phi (x)$ remains the usual one since it takes place in space ${\cal %
H.}$ Moreover, these are the observables and states that really matters
since they define $Z[\rho ].$ The observable $|O)$ is the generalized
version of eq. (\ref{6.7}) thus: 
\[
O=\sum_{N}\int dx_{1}\int dx_{2}...\int
dx_{N}O_{x_{1}x_{2}...x_{N}}^{(r)}|x_{1},x_{2},...x_{N})+ 
\]
\begin{equation}
\left. \sum_{N,\alpha _{i},i}\int dx_{1}\int dx_{2}...\int dx_{N-i}\frac{%
O_{N,x_{1}x_{2}...x_{N-i}}^{(\alpha _{i},s)}}{(n-4)^{\alpha _{i}}}|N,\alpha
_{i},x_{1},x_{2},...x_{N-i})\right]  \label{6.11}
\end{equation}
for all possible $N$ and all possible coincidence limits symbolized by $i$.
As before we can define an observable space

\begin{equation}
{\cal O=A}_{vhS}{\cal =}\widehat{{\cal A}}\oplus {\cal V}_{rS}\subset {\cal A%
}_S  \label{A10}
\end{equation}
where:

i.-The first term of the r.h.s. of eq. (\ref{6.11}) belongs to the space $%
{\cal V}_{rS}$ with basis \{$|x_{1},x_{2},...x_{N})\}$ and regular functions 
$O_{x_{1}x_{2}...x_{N}}^{(r)}.$

ii.-The second term of the r.h.s. of eq. (\ref{6.11}) belongs to the space $%
\widehat{{\cal A}},$ the algebra of the singularities of eq. (\ref{5.3})
with basis \{$|N,\alpha _{i},x_{1},x_{2},...x_{N-i})\}$ and regular
functions $O_{N,x_{1}x_{2}...x_{N-i}}^{(\alpha _{i},s)}.$ Then the singular
terms are like those of eq. (\ref{5.3}).

$(\rho |$ is the generalized version of the state $\rho (x_{1})\rho
(x_{2})...\rho (x_{N})|0\rangle \langle 0|.$ Then, if we repeat the
reasoning of eq. (\ref{6.8}), these generalized states would read: 
\[
\rho =\sum_{N}\int dx_{1}\int dx_{2}...\int dx_{N}\rho
_{x_{1}x_{2}...x_{N}}^{(r)}(x_{1},x_{2},...x_{N}|+ 
\]
\begin{equation}
\sum_{N,\alpha _{i},i}\int dx_{1}\int dx_{2}...\int dx_{N-i}\rho
_{N,x_{1}x_{2}...x_{N-i}}^{(\alpha _{i},s)}(N,\alpha
_{i},x_{1},x_{2},...x_{N-i}|  \label{6.12}
\end{equation}
As above we can defined the state space as:

\begin{equation}
{\cal O}^{\prime }{\cal =A}_{vhS}^{\prime }{\cal =}\widehat{{\cal A}}%
^{\prime }\oplus {\cal V}_{rS}^{\prime }\subset {\cal A}_S^{\prime }
\label{A11}
\end{equation}
where:

i.-The first term of the r.h.s. of eq. (\ref{6.12}) belongs to the space $%
{\cal V}_{rS}^{\prime }$ with basis \{$(x_{1},x_{2},...x_{N}|\}$ and regular
functions $\rho _{x_{1}x_{2}...x_{N}}^{(r)}.$

ii.-The second term of the r.h.s. of eq. (\ref{6.12}) belongs to the space $%
\widehat{{\cal A}},$ with basis \{$(N,\alpha _{i},x_{1},x_{2},...x_{N-i}|\}$
and regular functions $\rho _{N,x_{1}x_{2}...x_{N-i}}^{(\alpha _{i},s)}.$

Then: 
\[
(\rho |O)=\sum_N\int dx_1\int dx_2...\int dx_N\rho
_{x_1x_2...x_N}^{(r)}O_{x_1x_2...x_N}^{(r)}+ 
\]
\begin{equation}
\sum_{N,\alpha _i,i}\int dx_1\int dx_2...\int dx_{N-i}\rho
_{N,x_1x_2...x_{N-i}}^{(\alpha _i,s)}O_{N,x_1x_2...x_{N-i}}^{(\alpha
_i,s)}(n-4)^{-\alpha _i}  \label{6.13}
\end{equation}
which is only a mathematically well defined object when $n\rightarrow 4$ if
only the coordinates $\rho _{x_1x_2...x_N}^{(r)}$ and $%
O_{x_1x_2...x_N}^{(r)} $ do not vanish. But this is the case since either

i.- the physical observables really read 
\begin{equation}
O=\sum_N\left[ \int dx_1\int dx_2...\int
dx_NO_{x_1x_2...x_N}^{(r)}|x_1,x_2,...x_N)\right]  \label{6.14}
\end{equation}
since they have only the regular part (because they do not see the
singularities of the states, etc.) so they have no singular $(n-4)^{-\alpha
_i}$ terms or

ii.- the states really read

\begin{equation}
\rho =\sum_{N}\left[ \int dx_{1}\int dx_{2}...\int dx_{N}\rho
_{x_{1}x_{2}...x_{N}}^{(r)}(x_{1},x_{2},...x_{N}|\right]  \label{6.14'}
\end{equation}
since they have only the regular part (because they do not see the
singularities of the observables, etc.) so they have no singular $%
(n-4)^{-\alpha _{i}}$ terms. But here we have a better argument: they have
only regular part since the functions $\rho (x_{1})\rho (x_{2})...\rho
(x_{N})$ of eq. (\ref{6.10'''}) {\it are usually consider regular and with
no singularity.}

Therefore if we use the functional idea embodied in eq. (\ref{6.13}), or
better eq. (\ref{6.10'''}), and the regular state of eq. (\ref{6.14}) or
regular observables in (\ref{6.14'}) we just have: 
\begin{equation}
Z[\rho ]=\exp i\sum_{N}\int dx_{1}\int dx_{2}...\int
dx_{N}O_{x_{1}x_{2}...x_{N}}^{(r)}\rho _{x_{1}x_{2}...x_{N}}^{(r)}
\label{6.18}
\end{equation}
which is finite and the same happens with the $\partial /\partial \rho $
derivatives of $Z[\rho ].$ Thus the theory is finite.{\it \ So the theory
becomes finite just supposing that the physical observables} {\it \ are
regular (namely, just using as observables the real physical apparatuses in
our laboratory which give us {\bf finite measurements}) or the functions} $%
\rho (x_{1})\rho (x_{2})...\rho (x_{N})$ {\it are regular ({\bf which is the
usual }}{\bf supposition}) {\it and adopting the {\bf functional} approach
based in the ideas of the authors of papers }\cite{Haag}, \cite{Segal}, \cite
{Bogo}, and \cite{VanHowe}. In this way the substraction method is
justified. Instead if we use the naive usual formalism where all the
characters belong to Hilbert spaces and are multiplied using the ordinary 
{\it inner product} $Z[\rho ]$ will be singular and the theory must be
renormalized.

\section{Conclusion.}

Sometimes renormalization is considered as a {\it miracle} (\cite{Brown},
page 243, \cite{Ramond}, page 172). In fact: there is an infinite bare mass $%
m_{0}$ (which being infinite it can hardly be considered as ''bare''), and
an infinite counterterm, that plus the bare mass gives the finite physical
''dressed'' mass $m$ (which being finite is less dressed than the bare one);
there is an infinite bare coupling constant and a counterterm such
that,...etc. The substraction of all these infinities give (bingo!) the
right answer. This is a {\it pure miracle}\footnote{%
The author himself confesses that it was really difficult to understand and
to teach this miracle.}!

Now let us consider the same phenomenon according to the ideas of this
paper: We have chosen the simplest Lorentz invariant lagrangian $L,$
constructed using a scalar filed $\phi ,$ to base our theory. It is too much
to assume that $L$ would give us the right answers both for long and short
distances. In fact, it works remarkably well for long distances but it
behaves badly for short ones, since it produces short distance singularities
in the relevant N-points functions. So let us eliminate these singularities
and we will obtain the correct both short and long distance behavior. This
is the best we can do with lagrangian $L$ and the best we have until more
refined lagrangian will be invented (using perhaps superstrings, membranes,
etc.). Moreover, the singular structure is point-like and a pure
mathematical artifact, and therefore undetectable by the measurement
apparatuses, so it must be eliminated, in some way or other. So there is no
miracle in the finite nature of the theory and there is a logical
explanation of what really is going on. All these facts are embodied in the
rigorous mathematical structure of section VI.

Only a{\it \ minor miracle} remains. The numerical constant of some
(renormalizable) models are determined by a finite number of measurements,
while others (unrenormalizable) need an infinite number. Really it is a very
small miracle compared with the former one. We are used to deal with systems
that can be defined with a finite number parameters (e. g. mechanical
systems) while others have an infinite number (e.g.: the initial conditions
of classical electromagnetic fields or mechanical systems with an infinite
number of parameters like fluid with variable density or viscosity). Then
what really remains is a very big practical problem, how to work and solve
quantum field systems similar to the latter kind\footnote{%
Maybe superstrings or membranes are book keeping devices that allow us to
tame an infinite number of data as function $y=f(x)$ encompassing an
infinite number of data: the infinite relations between each value of
variable $x$ with each value of variable $y.$}. We do not propose a solution
but we believe that we have enlightened the real nature of the problem.

\section{Acknowledgment}

I would like to express my gratitude to my dear friends C. Bollini, Bocha
Giambiaggi, and A. Gonzalez Dominguez, that many years ago pointed out the
road I have followed in this paper and to I. Prigogine who several times
stressed the importance of singular structures in quantum field theory (even
with a completely different approach). This paper was partially supported by
grants PID-0150 of CONICET (Argentina National Research Council) and EX-198
of Buenos Aires University.

\end{document}